\documentclass[review]{elsarticle}

\usepackage{lineno,hyperref}
\modulolinenumbers[5]

\journal{Computer Speech and Language}
\usepackage{mathtools}

\usepackage{graphicx}
\usepackage{graphicx,epsf,float}
\usepackage{amsmath,amsfonts,amstext,amssymb,amsbsy,amsopn,amsthm,bm}
\usepackage{ifdraft,subcaption,color,units,psfrag}%,flushend}
\usepackage[nolist]{acronym}
\usepackage{hyperref}

\hypersetup{
    colorlinks=true,
    citecolor=blue,
    linkcolor=black,
    linkbordercolor={1 1 1}}
\usepackage{balance}

\usepackage{booktabs}
\usepackage{tikz}
\usepackage{pgfplots}
\pgfplotsset{compat=newest}
\usepackage{steinmetz}

\usetikzlibrary{arrows, decorations.markings}
\usepackage{graphicx}
\def\infinity{\rotatebox{90}{8}}
\usepgfplotslibrary{groupplots}
\usepackage[width=18cm]{geometry}
\usepackage{pgfplots}
\usepgfplotslibrary{groupplots}
\usetikzlibrary{pgfplots.groupplots}
\usetikzlibrary{plotmarks}
\usetikzlibrary{calc}
\usepgfplotslibrary{external}
\usepackage{amssymb}% http://ctan.org/pkg/amssymb
\usepackage{pifont}
\newcommand{\cmark}{\ding{51}}%
\newcommand{\xmark}{\ding{55}}%

\usepackage{breqn}
\usepackage{numprint}
\usepackage{booktabs}
\usepackage{graphicx}
\usepackage{color}
\usepackage{calc}
\usepackage{pgfplots}
\usepgfplotslibrary{groupplots}
\usepackage{amsmath}

\biboptions{sort&compress}

\begin{acronym}
\acro{DOA}[DOA]{direction-of-arrival}
\acro{CCE}[CCE]{categorical cross-entropy}
\acro{E2E}[E2E]{end-to-end}
\acro{STFT}[STFT]{short-time Fourier transform}
\acro{SOI}[SOI]{source-of-interest}
\acro{DNN}[DNN]{deep neural network}
\acro{CNN}[CNN]{convolutional neural network}
\acro{RIR}[RIR]{room impulse response}
\acro{FFNN}[FFNN]{feed-forward neural network}
\acro{SPS}[SPS]{spatial pseudo spectrum}
\acro{AE}[AE]{absolute error}
\acro{flops}[FLOPs]{floating-point-operations}
\end{acronym}

\begin{document}

\begin{frontmatter}

\title{ Signal-Aware Direction-of-Arrival Estimation Using Attention Mechanisms}

\author[mymainaddress]{Wolfgang Mack\corref{mycorrespondingauthor}}
\cortext[mycorrespondingauthor]{Corresponding author}
\ead{wolfgang.mack@audiolabs-erlangen.de}
\author[mysecondaryaddress]{Julian Wechsler}
\ead{julian.wechsler@fau.de}
\author[mymainaddress]{Emanu\"el A. P. Habets}
\ead{emanuel.habets@audiolabs-erlangen.de}

\address[mymainaddress]{International Audio Laboratories Erlangen (a joint institution of the Friedrich-Alexander-University Erlangen-Nuremberg (FAU) and Fraunhofer IIS), Am Wolfsmantel 33, 91058 Erlangen, Germany.}
\address[mysecondaryaddress]{Friedrich-Alexander-University Erlangen-Nuremberg, Schloßplatz 4, 91054 Erlangen, Germany.}
\begin{abstract}
The \ac{DOA} of sound sources is an essential acoustic parameter used, e.g., for multi-channel speech enhancement or source tracking. Complex acoustic scenarios consisting of sources-of-interest, interfering sources, reverberation, and noise make the estimation of the \ac{DOA}s corresponding to the sources-of-interest a challenging task. Recently proposed attention mechanisms allow \ac{DOA} estimators to focus on the sources-of-interest and disregard interference and noise, i.e., they are signal-aware. The attention is typically obtained by a \ac{DNN} from a \ac{STFT} based representation of a single microphone signal. Subsequently, attention has been applied as binary or ratio weighting to \ac{STFT}-based microphone signal representations to reduce the impact of frequency bins dominated by noise, interference, or reverberation. The impact of attention on \ac{DOA} estimators and different training strategies for attention and \ac{DOA} \ac{DNN}s are not yet studied in depth. In this paper, we evaluate systems consisting of different \ac{DNN}s and signal processing-based methods for \ac{DOA} estimation when attention is applied. Additionally, we propose training strategies for attention-based \ac{DOA} estimation optimized via a \ac{DOA} objective, i.e., end-to-end. The evaluation of the proposed and the baseline systems is performed using data generated with simulated and measured room impulse responses under various acoustic conditions, like reverberation times, noise, and source array distances. The best-performing systems are also evaluated using measured data. Our experiments show that  \ac{DNN}s used for \ac{DOA} estimation are biased to the spectral source characteristics and the spectral attention distribution used during training (e.g., spectrally flat/sparse). We also show that this bias in the  \ac{DOA} estimator can be avoided if signal-processing methods are used in combination with attention.  Overall,  \ac{DOA} estimation using attention in combination with signal-processing methods exhibits a far lower computational complexity than a fully \ac{DNN}-based system; however, it yields comparable results.
\end{abstract}
\acresetall

\end{frontmatter}

\section{Introduction}
The sound emitted by a point source in an enclosed space spreads spherically and gets reflected by walls and other obstacles. Typically, the non-reflected sound is referred to as direct, whereas reflections are referred to as reverberation. Additional interfering sources like a ventilator, or background noise, e.g., from a nearby road, add further complexity to the sound field and severely degrade humans' and machines' ability to localize sources or understand speech. When such a sound field is captured with an array of microphones, acoustic signal-processing techniques can be used to increase the speech intelligibility, e.g., with beamformers \cite{ BENESTY2014, Benesty2011,gannot2017consolidated, Habets2009b, Souden2010b}, or to track sources \cite{lollmann2018locata},  e.g., with acoustic simultaneous localization and mapping \cite{evers2018acoustic}. These techniques often require estimates of the \ac{DOA} of the sound sources. For some applications, only the \ac{DOA} of the \ac{SOI} is desired or required.  For example, consider two concurrently active sources: \ac{SOI} and interference. Conventional \ac{DOA} estimators yield the \ac{DOA} of both sources. Typically, it is unclear which of both \ac{DOA}s corresponds to the \ac{SOI} and which to the interfering source.  A correct \ac{DOA} assignment to the \ac{SOI} is crucial for beamformers, as a wrong assignment leads to an attenuation of the \ac{SOI}.

Estimation methods for the \ac{DOA} have been investigated thoroughly in the literature (e.g., \cite{ferguson2019introduction, Chen2010a, Tuncer2009}).  Typically,  \ac{DOA} estimators exploit spatial features present in the microphone signals due to a level difference and a time difference of arrival (TDOA) of the source signal between the individual microphones.  For far-field scenarios and omnidirectional microphones,  the level difference is usually minimal and can be neglected. In the frequency domain, the TDOA translates to frequency-band specific phase-differences between the microphones. Here, the spatial features can be exploited per frequency band (narrowband) or from all frequency bands (broadband).  Signal processing-based methods to estimate the \ac{DOA} from these features can be based on the inter-microphone cross-correlations \cite{MCCC1, MCCC2,Carter1973},  beamformers \cite{mvdrsteer,SRP, Johansson2004} or subspaces \cite{MUSIC, Johansson2004, MLDOA,Roy1989, Bermudez2009, Teutsch2005a, jo2018direction}.  The inter-microphone cross-correlations, like the generalized-cross-correlation \cite{Knapp1976} with maximum likelihood or phase transform (PHAT) \cite{Carter1973} weighting exploit the \ac{DOA} and frequency dependency of the inter-microphone phase-differences to estimate the \ac{DOA}.  In \cite{Leastsquaredoa},  a least square approach is used to minimize the phase differences obtained from measurements and an estimated \ac{DOA}.  Beamforming techniques like steered-response power, e.g., with PHAT weighting \cite{SRP, Johansson2004} (SRP-P), or a steered minimum-variance distortionless response beamformer \cite{mvdrsteer} sample the \ac{DOA} space by steering in pre-defined directions. The \ac{DOA} is subsequently estimated by maximum picking. An alternative is null-steering \cite{AdaptiveNullsteering}, where the idea is similar, but the \ac{DOA} is obtained by minimum picking. Alternatively, the \ac{DOA} can be estimated using subspace-based methods like MUltiple SIgnal Classification (MUSIC) \cite{MUSIC, Johansson2004, MLDOA}, based on noise subspaces, or estimation of signal parameters via rotational invariant techniques \cite{Roy1989, Bermudez2009, Teutsch2005a, jo2018direction}, based on sub-arrays.  In \cite{Goldsthein2017}, the authors exploit reflection patterns using semi-supervised manifold learning with a distributed microphone array to localize a single source. 

 Also, deep-learning techniques have been used for \ac{DOA} estimation \cite{perotin2018crnn,Perotin2019,hirvonen2015,MA2017,Vesperini2016,Takeda2016,Xiao2015,Takeda2016a,Yalta2017,Adavanne2018,He2018,Chakrabarty2017a,TNT2020, Chakrabarty2017b,sharath2018,9357962,sharath2019,zhang2019robust, chakrabarty2019multi,wang2018target,Sivasankaran2018, Mack2020SigAware, kuccuk2019deep, chakrabarty2019multiscale, Guerra2020}.  Typically, deep-learning methods for \ac{DOA} estimation are computationally more complex than signal-processing methods and require retraining or architecture modifications if fundamental parameters like the number of microphones or the array architecture change.  When using deep learning for \ac{DOA} estimation,  a \ac{DNN} learns to map a feature representation of the microphone signals to the \ac{DOA}.  This enables matching the trained \ac{DNN} via the training data to specific scenarios. The estimated  \ac{DOA} on the \ac{DNN} output can be represented in a classification manner, where a class activity symbolizes an active source from the corresponding direction,  or a regression manner, where a single variable represents the \ac{DOA} (e.g., an angle).  According to \cite{Perotin2019}, both representations yield comparable results such that the output representation of the \ac{DOA} is a design choice.   Some of the \ac{DNN}s for \ac{DOA} estimation (referred to as DDNNs) are trained with directional noise signals (e.g., \cite{Chakrabarty2017b, chakrabarty2019multi,  Huebner2021}) as this allows to generate an infinite amount of simulated training data.  Conceptually,  training with noise implies that the DDNN learns spatial and no spectral source characteristics, as expected of a \ac{DOA} estimator.  Very recently,  \cite{vargas2021improved} showed that training with speech improves the localization of speech sources compared to training with noise, although the DDNN was only provided with the \ac{STFT} phases of the microphone signals and not the respective magnitudes. The DDNN, consequently, is biased towards speech sources if trained with speech or towards spectrally white sources when trained with spectrally white noise.  In comparison, signal-processing methods for \ac{DOA} estimation do not exhibit such a bias towards spectral source characteristics. 

In the so-called sound event localization and detection (SELD) task \cite{hirvonen2015,sharath2018,sharath2019},  the ability of \ac{DNN}s to be tailored to specific (spectral) source characteristics for \ac{DOA} estimation is exploited to localize specific sources and their activity, only. In SELD, a \ac{DNN} is used to detect a sound event and the respective \ac{DOA}.  In \cite{hirvonen2015,sharath2018,sharath2019}, the sources-of-interest have to be defined during training the DDNN. For each \ac{SOI} class, the DDNN has outputs for the respective source activity and the \ac{DOA}. This approach requires retraining the DDNN if the \ac{SOI} changes. Additionally, the number of DDNN outputs has to be increased each time the number of \ac{SOI} classes increases, which could introduce scaling problems for a high number of \ac{SOI} classes, or changing classes. An example of many changing classes is the localization of multiple speakers,  where each class represents a specific speaker.  In \cite{hirvonen2015,sharath2018,sharath2019}, each \ac{SOI} speaker had to be defined during training and treated as an individual class on the DDNN output. Changing \ac{SOI} speakers would require retraining the DDNN, which is impractical for many applications. 

In SELD, a single \ac{DNN} is designed to detect the activity of sources-of-interest and determine their \ac{DOA}s. Alternatively,  these two tasks can be separated via the concept of attention \cite{zhang2019robust,wang2018target, Sivasankaran2018, Mack2020SigAware,ZWang2019, Pretilae2017, wang2018target,Xu2017WSPS}.  Systems using attention for \ac{DOA} estimation often consist of a \ac{DOA} module, which estimates the \ac{DOA} (e.g., a DDNN or a signal-processing method)  and an attention module, which provides the attention that is used in the \ac{DOA} module to focus on the \ac{SOI} and disregard interference, reverberation, and noise.  
If the \ac{SOI} selection is performed independently of the \ac{DOA} module (e.g., if the \ac{DOA} module is a signal-processing method),  no \ac{DOA} module change (architecture change,  retraining) is required for a changing or an increasing number of \ac{SOI} classes as in SELD \cite{hirvonen2015,sharath2018,sharath2019}.  Only the attention has to be modified.  Attention can be estimated from spectral \cite{zhang2019robust,wang2018target, Sivasankaran2018, Mack2020SigAware} or spatial \cite{subramanian2020directional} features and can be implemented as a weighting applied to a feature representation of the microphone signals in or before the \ac{DOA} module. The weighting concept of attention is similar to the time-frequency masking concept for single-channel source extraction/separation/enhancement (e.g., \cite{Williamson2016, Williamson2017, Hershey2016, Chen2017, Isik2016, Wang2014, Yu2017, Luo2017, wang2019deep,8664086}), which allows adopting the concepts of this highly investigated field to compute attention to enable signal-aware \ac{DOA} estimation.  For example, a promising direction is to adopt techniques from universal sound source separation for attention-based multi-source \ac{DOA} estimation \cite{Tzinis2020, Kavalerov2019}. Consequently, attention provides additional flexibility. 

Typically, the attention module is implemented in the form of a \ac{DNN} (referred to as ADNN) for both fully \ac{DNN}-based systems and hybrid systems, where a signal-processing method is used as \ac{DOA} module. In hybrid systems \cite{ZWang2019, Pretilae2017, wang2018target,Xu2017WSPS},  the ADNN estimates a time-frequency mask and was optimized using the ideal binary mask \cite{Xu2017WSPS, wang2018target}, the Wiener filter \cite{Pretilae2017},  or the phase-sensitive mask \cite{ZWang2019} as the target.  That way,  feature representations of the input signals are modified to be dominated by the \ac{SOI}. Subsequently,  MUSIC \cite{Xu2017WSPS}, SRP-P \cite{ZWang2019, Pretilae2017}, or a complex Watson mixture model  \cite{wang2018target} were used to estimate the \ac{DOA} of the \ac{SOI}.  The \ac{SOI}, thereby, was exclusively defined as a speech source. In \cite{wang2018target},  a multi-speaker environment was considered where the \ac{SOI} was defined in a speakerbeam \cite{speakerbeam} like manner using a reference audio snippet of the respective \ac{SOI} (speaker) processed in the ADNN.  In \cite{ZWang2019, Pretilae2017,Xu2017WSPS}, the environment consisted of a single speech source (the \ac{SOI}) and non-speech interference or babble noise.  

In fully \ac{DNN}-based systems \cite{Mack2020SigAware,zhang2019robust, Sivasankaran2018}, the ADNN estimates attention for a DDNN to enable signal-aware \ac{DOA} estimation.  The DDNNs typically consist of a \ac{CNN} to extract features from the microphone signals and a subsequent \ac{FFNN} to map the features to the \ac{DOA}.  Attention has been applied either before \cite{Mack2020SigAware,Sivasankaran2018,zhang2019robust} or after the feature extracting \ac{CNN} \cite{Mack2020SigAware, subramanian2020directional}.  In our preliminary work  \cite{Mack2020SigAware}, we showed that attention application after the \ac{CNN} outperforms attention application at the input. Training the ADNN has been done using masking-based \cite{Sivasankaran2018, Mack2020SigAware,zhang2019robust} or \ac{E2E} using \ac{DOA}-based \cite{zhang2019robust} objectives.  For low-SNR scenarios and unmatched training-test conditions, \ac{E2E} training performed best, whereas, for high SNR scenarios, attention degraded the performance slightly compared to the attention-free scenario \cite{zhang2019robust}.  As for the hybrid systems, the focus of the fully \ac{DNN}-based methods is on localizing speech sources.  In \cite{Sivasankaran2018}, the \ac{SOI} is a specific speaker defined by a spoken keyword in a multi-speaker environment,  whereas in \cite{Mack2020SigAware, zhang2019robust},  there is only a single speaker and non-speech interference.  

To this point, it is not clear how hybrid systems perform in comparison to fully \ac{DNN}-based systems, although hybrid systems typically exhibit a far lower computational complexity. Additionally,  the effect of attention on DDNNs is not yet investigated thoroughly. To investigate the effect of attention on DDNNs,  we evaluate different DDNN architectures, input features,   attention-application methods,  and training data simulation methods and compare their performance on data simulated using measured room impulse responses (RIRs).  In particular, we investigate the importance of spectral context for DDNNs (see Sections~\ref{CHAP:3Narrowband}, \ref{CHAP:E2ETRAININGNB}, \ref{SUBSEC:PEV1}),  whether DDNNs are biased towards a specific attention distribution via training (see Section~\ref{SUBSEC:PEV1}) and whether the DDNN architecture can be reduced significantly dependent on the input features/architecture (see Sections~\ref{CHAP:3.1},  \ref{SUBSEC:PEV1}).   Subsequently, we evaluate different attention-application and DDNN/ADNN training methods in Section~\ref{CHAP:E2EEVALUATION}.  Finally,  the best performing fully \ac{DNN}-based system is compared to different hybrid systems for signal-aware \ac{DOA} estimation of a single speech source in the presence of directional interference and noise.   Experiments with a fine-grained \ac{DOA} resolution ($\approx 1^\circ$) and using measured data are conducted in Section~\ref{CHAP:FG} and Section~\ref{CHAP:MEAS}, respectively. 

The main contributions can be summarized as follows: (I) Exhaustive evaluation and comparison of existing and proposed methods for signal-aware \ac{DOA} estimation using simulated and measured data in Section~\ref{CHAP:5}.  (II) The proposition of novel training methods for fully \ac{DNN}-based and hybrid systems in Section~\ref{CHAP:3Attention}.  In particular, we propose: a) An \ac{E2E} training method when attention is applied in the DDNN - differences to \cite{zhang2019robust} are explained in Section~\ref{SUBSEC:AttAppl} and to \cite{subramanian2020directional} in Section~\ref{CHAP:E2ETFMPM}; b) A narrowband DDNN; c)  Training an ADNN in combination with SRP-P with a \ac{DOA}-related loss (state-of-the-art is based on masking/enhancement-related losses \cite{ZWang2019, Pretilae2017, wang2018target,Xu2017WSPS}) (III) Evaluation of the role of attention and spectral context for DDNNs.  The work presented here builds upon our work \cite{Mack2020SigAware} presented at ICASSP~2020.

The remainder of the paper is organized as follows. In Section~\ref{CHAP:2}, we introduce a signal model and introduce our preliminary work \cite{Mack2020SigAware} and baselines for hybrid \cite{ZWang2019} and fully \ac{DNN}-based \cite{zhang2019robust} signal-aware \ac{DOA} estimation. Subsequently, we introduce the proposed modifications to the \ac{DNN} architecture from \cite{Mack2020SigAware} and the proposed training methods for hybrid and fully \ac{DNN}-based systems.  In Section~\ref{CHAP:4}, we describe the data sets used for evaluation and training.  Finally, in Section~\ref{CHAP:5},  we evaluate the proposed and baseline systems using measured and simulated data.

\section{Fundamentals}
\label{CHAP:2}
\subsection{Problem Formulation}
\label{CHAP:2.1}
\begin{figure}
\centering
\includegraphics{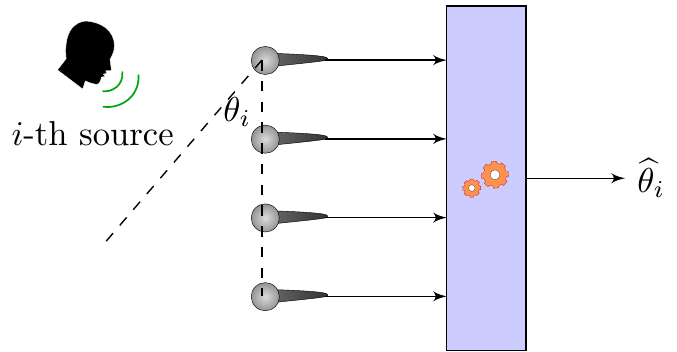}
\caption{Scheme for \ac{DOA} ($\theta_i$) estimation of the $i$-th source. }
\label{FIG:SIGNALMODEL}
\end{figure}
We assume a uniform-linear microphone array (ULA) with $Q$ microphones with microphone index $q \in \{1,\ldots,Q\}$ positioned in a reverberant room with several directional sound sources. We define the microphone signals in the \ac{STFT} domain as $Y\in\mathbb{C}^{Q \times K \times N}$, where $\mathbb{C}$ specifies the complex domain,  $K$ specifies the number of frequencies of the one-sided \ac{STFT} spectrum with frequency-index $k \in \{1,...,K\}$ and $N$ specifies the number of time-frames with time-frame index  $n \in \{1,...,N\}$.  The microphone signals can be modelled as a superposition of representations of the source signals $X_i \in\mathbb{C}^{Q \times K \times N}$ with source index $i$  and spatio-temporally white microphone self-noise $V \in\mathbb{C}^{Q \times K \times N}$, i.e., 
\begin{equation}
Y = \sum_{i=1}^{I} X_i + V, 
\end{equation}
where $I \in \mathbb{N}$ is the number of sources. The source  $X_i $ can be split into 
a direct component $X_i ^{\textrm{d}}\in\mathbb{C}^{Q \times K \times N}$ that reaches the microphones without being reflected (e.g., from walls,  or other obstacles) and a reverberant component $X_i^{\textrm{r}}\in\mathbb{C}^{Q \times K \times N}$, i.e., 
\begin{equation}
X_i = X_i^{\textrm{d}} + X_i^{\textrm{r}}.
\end{equation}
For a ULA, the \ac{DOA} of the $i$-th source is defined as the angle $\theta_{i} \in [0,180]$ which specifies the direction of the $i$-th source to the microphone array as shown in Figure~\ref{FIG:SIGNALMODEL}.  From Figure~\ref{FIG:SIGNALMODEL}, it can be inferred that this information is embedded in $X_i^{\textrm{d}}$.  Other sources, reverberation,  or noise, consequently complicate the \ac{DOA} estimation of the $i$-th source. 

In a typical \ac{DOA} estimation context, the objective is to estimate from $Y$ the \ac{DOA} of all $I$ sources.  In signal-aware \ac{DOA} estimation, the aim is to estimate from $Y$ only the \ac{DOA}s of the sources-of-interest, which are in a subset of all $I$ sources.  The definition of sources-of-interest,  thereby, is user and application-defined. To reduce the impact of reverberation, noise, and interference on the \ac{DOA} estimate and enable signal-aware \ac{DOA} estimation, attention in the form of a weighting mask $M \in [0,1]^{K \times N}$ can be applied to $Y$ or a feature representation of it.  In \ac{STFT} domain, the smallest unit to estimate the \ac{DOA} from is a single \ac{STFT}-bin $Y[:,k,n]$. The largest unit is the whole signal $Y[:,:,:]$. Strong interference or pauses of the \ac{SOI} can deteriorate to the final \ac{DOA} estimate. Via weighting with $M$, the influence of \ac{STFT}-bins that degrade the \ac{DOA} estimation performance of the sources-of-interest can be reduced. 

If $M$ attends to multiple directional sound sources, there is no assignment of the estimated \ac{DOA}s to the respective sources-of-interest.  For multiple sources-of-interest, the weighting $M$ can be constructed such that all sources-of-interest are attended to simultaneously. Alternatively, different sound classes can be defined, where each class, for example,  represents a single speaker or a collection of directional sounds.  In this case,  a weighting  $M_u$ can be constructed to attend to a source belonging to the $u$-th class. This process can be repeated for all sources-of-interest to estimate their respective \ac{DOA}s. For example, consider two active sources-of-interest, a speaker and a loudspeaker playing music.  When both sources are attended to simultaneously with a single mask,  two \ac{DOA}s are obtained without an assignment to speech and music.  With $M_1$ and $M_2$, each time one \ac{DOA} is obtained, where the assignment of $M_1$ and $M_2$ to speech and music, respectively, allows assigning the respective \ac{DOA}s to speech and music in the same way.  This has the additional advantage that two \ac{DOA}s are obtained even if the sources come from the same direction.  Using multiple masks enables multi-source \ac{DOA} estimation via single-source \ac{DOA} estimation.  Consequently,  we restrict the experiments to a single \ac{SOI} of type speech without loss of generality. 
\subsection{Fully \ac{DNN}-Based Signal-Aware \ac{DOA} Estimation}
\label{CHAP:2.2}
\begin{figure*}
\centering
\includegraphics{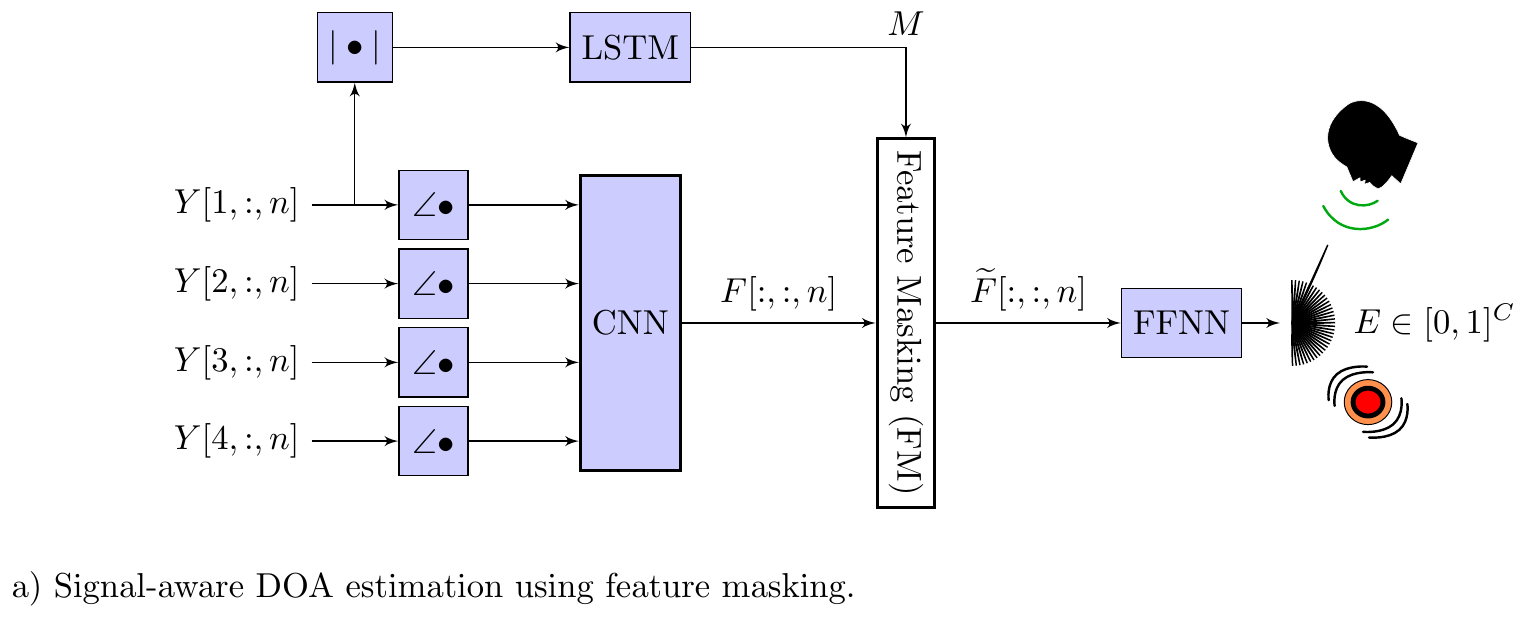}

\vspace{1.5cm}

\includegraphics{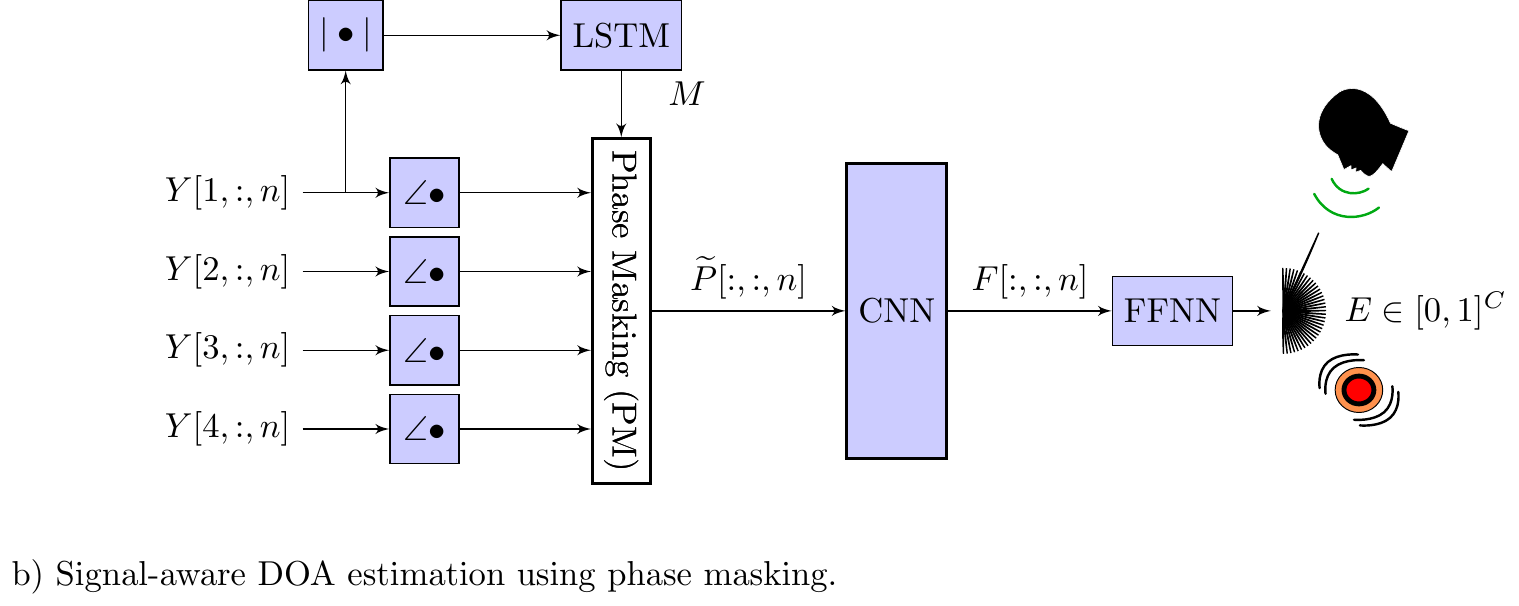}
\caption{Scheme of the \ac{DNN} system from our preliminary work \cite{Mack2020SigAware}. $|\bullet|$ symbolizes magnitude extraction of the input and $\angle \bullet$ phase extraction of the respective input. A DDNN based on a \ac{CNN} and  a \ac{FFNN} maps the \ac{STFT} phases frame-wise to 37 \ac{DOA} classes. Attention is included in a) via feature masking [see (\ref{EQU:B-FM})] and alternatively in b) via phase masking [see (\ref{EQU:B-PM})] such that only the \ac{SOI}, here a speech source, is localized. The attention is computed from the magnitude spectrum of one microphone using a long short-term memory neural network (LSTM), the ADNN. }
\label{Fig:Baseline}
\end{figure*}
In this section, we review our recently proposed \ac{DNN} for signal-aware \ac{DOA} estimation \cite{Mack2020SigAware}. The architecture is depicted in Figure~\ref{Fig:Baseline}.
\subsubsection{Architecture and Training}
The DDNN \cite{chakrabarty2019multi} consists of two parts, a \ac{CNN} and a \ac{FFNN}. The \ac{CNN} consists of $Q-1$ convolutional layers with $64$ filters of shape $(2,1)$ (inter-microphone application), each, stride~$1$, padding~$0$, and ReLU activation after each layer. The \ac{FFNN} consists of 3 layers with ReLU activation and a sigmoid output activation with shapes $(257\cdot 64, 512)$, $(512, 512)$, $(512, C)$, where $C$ represents a discrete representation of the \ac{DOA} space corresponding to angles $\widetilde{\theta} \in \mathbb{R}^C$, where here $C=37$, and $\widetilde{\theta}[c] = (c-1)\cdot \frac{180}{C-1}^\circ$, with $c \in \{1, \ldots, C\}$. We denote the discrete \ac{DOA} representation for $N$ consecutive time-frames as $E \in [0,1]^{C\times N} $. The input of the DDNN consists of the microphone phases of a single time-frame $n$, $P[:,:,n] \in [-\pi,\pi]^{Q\times K}$. We denote the features after the \ac{CNN} for $N$ separately processed time-frames as $F \in \mathbb{R}^{K\times 64\times N}$, where 64 is the total number of \ac{CNN} filters, and $N$ specifies the number of processed time-frames. Consequently, the \ac{CNN} filters extract inter-microphone but not inter-frequency information from $P[:,:,n]$.   Finally, the \ac{FFNN} maps $F[:,:,n] $ to $E[:,n]$, where values close to $1$ specify source activity in the respective direction in time-frame $n$ \cite{chakrabarty2019multi}, and zero specifies no activity. The DDNN is trained with the \ac{CCE} loss on a time-frame basis. Training data was simulated with noise and simulated RIRs as in \cite{Chakrabarty2017a}.

For evaluation, the time-frame estimates in  $E$ can be combined by averaging, i.e.,
\begin{equation}
\bar{E}[c]= \frac{1}{N}\cdot\sum_{n=1}^{N}E[c,n] \in [0,1],
\end{equation}
to obtain a global estimate. The estimated \ac{DOA} of the \ac{SOI} is $\widetilde{\theta}[\textrm{argmax}\{\bar{E}\}] $ where ``$\textrm{argmax}$'' returns the index of the maximum.

The ADNN maps the magnitude representation of a single microphone $|Y[1,:,:]|$ to a ratio mask $M_{r} \in [0,1]^{K\times N}$ of equal size for the speech source. The ADNN consist of a bidirectional long short-term memory neural network (BLSTM) \cite{Hochreiter1997} with 3 layers and $1200$ neurons per layer. The output layer is a feed-forward layer with sigmoid activation. The ADNN is trained for a single-channel speech enhancement objective, to minimize the mean-squared error (MSE) between the direct signal of the \ac{SOI} and the respective estimate at the first microphone,
\begin{equation}
  \text{MSE} = \frac{1}{K\cdot N}\cdot \sum_n \sum_k (|X^{\textrm{d}}_{1}[1,k,n]| - |M_{r}[k,n]\cdot Y[1,k,n]|)^2,
\label{EQU:MSESPEECHENHANCEMENT}
\end{equation}
where only Source~1 is a speech source and all other sources are non-speech.
\subsubsection{Attention Application}
\label{SUBSEC:AttAppl}
In \cite{Mack2020SigAware}, we proposed to compute binary attention $ M_{b}[k,n]$ for the speech source via thresholding, i.e.,
\begin{equation}
    M_{\text{b}}[k,n]= 
\begin{dcases}
    0 & \text{if } M_{\text{r}}[k,n] < v_{\text{thr}}; \\
    1            & \text{otherwise,}
\end{dcases}
\end{equation}
where the threshold $v_{\text{thr}}\in [0,1]$. Subsequently, $ M_{\text{b}}$ can be applied to the DDNN via two methods. First, via binary phase-masking (B-PM) by directly modifying the DDNN input $P$, such that
\begin{equation}
  \label{EQU:B-PM}
\widetilde{P}[q,k,n] =  M_{\text{b}}[k,n] \cdot P[q,k,n]  + (1- M_{\text{b}}[k,n])\cdot \nu [q,k,n],
\end{equation}
where $\nu[q,k,n] \in [-\pi,\pi]$ is a uniformly distributed random variable and $\widetilde{P}[:,:,n]$ is the new DDNN input. Note that $\nu[q,k,n]$ is used to avoid all-zero phase inputs in a frequency band as this would correspond to a \ac{DOA} of $90^{\circ}$. For robust \ac{DOA} estimation of a single source in the presence of reverberation and noise, the authors in \cite{zhang2019robust} proposed a similar phase masking approach as in (\ref{EQU:B-PM}), with $v_q = 0$ and ratio attention $M_{\text{r}}[k,n] \in [0,1] $ instead of $M_{\text{b}}[k,n]$. We refer to phase masking with $M_{\text{r}}$ and $v_q = 0$ using our \ac{DNN} architecture as ratio phase-masking without randomization (referred to as R-PM*) and compare it to B-PM and ratio phase-masking with randomization, which is described in Section~\ref{CHAP:E2ETFMPM}.

Secondly, attention can be applied via binary feature-masking (B-FM) by zeroing selected features after the \ac{CNN}, i.e.,
\begin{equation}
  \label{EQU:B-FM}
  \widetilde{F}[k,:,n] = F[k,:,n] \cdot  M_{\text{b}}[k,n],
\end{equation}
where $ F[:,:,n]$ are the features after the \ac{CNN}, and $\widetilde{F}[:,:,n]$ are masked features (see Figure~
\ref{Fig:Baseline}) which are fed in the \ac{FFNN} instead of $ F[:,:,n]$ when B-FM is used.  The same binary attention value, thereby, is used for all 64 features in a specific time-frequency bin. Please note that either B-PM or B-FM is applied in previous works and that an application of both would be equal to B-FM \cite{Mack2020SigAware}.
\subsection{Baseline System: Hybrid Signal-Aware \ac{DOA} Estimation}
\label{CHAP:2.3}
 In \cite{ZWang2019, Pretilae2017}, the authors proposed a steered-response power with modified PHAT weighting (SRP-MP) to perform signal-aware \ac{DOA} estimation. Attention, thereby, is used to modify the PHAT weighting. In \cite{Pretilae2017}, attention is computed from averaged microphone features using an ADNN trained for a Wiener filter objective. In \cite{ZWang2019}, attention is computed from each microphone separately and is  trained using a variant of the phase-sensitive mask (PSM) for the speech source as the target, i.e.,
 \begin{equation}
  \text{PSM}[q,k,n] =  \max \left\{ 0 ,\sqrt{ \frac{|X^{\textrm{d}}_{1}[q,k,n] |^2}{|X^{\textrm{d}}_{1}[q,k,n] |^2 + |Y[q,k,n] - X^{\textrm{d}}_{1}[q,k,n] |^2} } \cdot \cos ( \angle X^{\textrm{d}}_{1}[q,k,n]  - P[	q,k,n] ) \right\}, 
  \label{EQU:PSM}
 \end{equation}
where $\angle X^{\textrm{d}}_{1}[q,k,n]  $ provides the phase of $ X^{\textrm{d}}_{1}[q,k,n]  $. In SRP-P and SRP-MP, the microphone signals are transformed in the \ac{STFT} domain, and each time-frequency bin is normalized with its magnitude (PHAT weighting).  The PHAT weighting can be denoted as 
\begin{equation}
    \text{W}[q,k,n]= 
\begin{dcases}
    \frac{1}{|Y[q,k,n]|} & \text{if } |Y[q,k,n]| > \epsilon; \\
    \epsilon           & \text{otherwise,}
\end{dcases}
\label{EQU:PHATWEIGHTING}
\end{equation}
with the small constant $\epsilon \in \mathbb{R}^+$.   Subsequently, the power of the normalized spectrum coming from different sampled directions is computed.  The \ac{DOA} is obtained by picking the direction with the maximum power.

For signal-aware \ac{DOA} estimation,  the PHAT weighting can be modified by applying the mask $M_{\text{r}}$ to it, i.e.,
\begin{equation}
\widetilde{\text{W}}[q,k,n]=\text{W}[q,k,n]\cdot M_{\text{r}}[k,n],
\label{EQU:MPHAT}
\end{equation}
 to obtain a new weighting function. In \cite{Pretilae2017}, the same $M_{\text{r}}$ is used for all microphones, whereas in \cite{ZWang2019}, the mask is channel dependent.  We refer to the application of the SRP-P algorithm with weighting  $\widetilde{\text{W}}$ as SRP-MP\footnote{Derivation of the equality of PHAT weighting and (\ref{EQU:PHATWEIGHTING}) can be found in \cite{SRP}.}. Subsequently, the weighting is applied to the microphone signals and the matrix $\Phi \in \mathbb{C}^{K\times N \times Q \times Q}$ is computed, i.e.,
\begin{equation}
\Phi[k,n,q_1,q_2] = Y[q_1,k,n]\cdot \widetilde{\text{W}}[q_1,k,n]\cdot\widetilde{\text{W}}[q_2,k,n]\cdot  Y^*[q_2,k,n],
\label{Equ:PHIYY}
\end{equation}
where $^*$ denotes the complex conjugate. As in SRP-P \cite{SRP}, the \ac{DOA} space is sampled assuming a far-field model and relative transfer functions $D \in \mathbb{C}^{C\times K\times Q} $ w.r.t. the first microphone, where a single element is denoted as
\begin{equation}
  D[c,k,q] = \mathrm{e}^{\frac{-j\cdot2\cdot\pi\cdot k\cdot f_s\cdot\cos(\widetilde{\theta}[c])\cdot d_{q}}{c_s\cdot K}},
\end{equation}
where $\mathrm{e}$ is Euler's number, $c_s$ is the speed of sound, $f_s$ is the sampling frequency, $j=\sqrt{-1}$ is the complex unit, and $d_{q}$ denotes the distance between the first and the $q$-th microphone. Finally,  SRP-MP steered to all elements in $\widetilde{\theta}$ is obtained via
\begin{equation}
\text{SRP-MP}[c] (\Phi) =
\sum_{n=1}^{N} \sum_{k=1}^{K}  \mathop{\sum_{q=1}^{Q} \sum_{j =q +1}^{Q}}\left(\frac{ 2\cdot \textrm{Real}\{ \text{D}[c,k,q]\cdot \Phi[k,n,q,j]\cdot \text{D}^*[c,k,j]\}}{N\cdot K\cdot (Q-1)\cdot (Q-1)}\right),
\label{Equ:SRP}
\end{equation}
where ``Real''  provides the real part, only.
Finally, $\bar{E}_{\text{SRP-MP}}$ is obtained by normalizing $\text{SRP-MP}$, i.e.,
\begin{equation}
\bar{E}_{\text{SRP-MP}}[c] = \frac{\text{SRP-MP}[c]}{\text{max}_{1\leq u\leq C}(\text{SRP-MP}[u])}.
\label{EQU:SRPnormalize}
\end{equation}
The same normalization procedure can be applied to SRP-P, which results in $\bar{E}_{\text{SRP-P}}$. The estimated \ac{DOA} $\widehat{\theta}$ of the \ac{SOI} is $\widetilde{\theta}[\textrm{argmax}\{\bar{E}\}] $ where ``$\textrm{argmax}$'' returns the index of the maximum.  Note that the SRP-MP approach differs from SRP-P \cite{SRP} only in the application of $M_{\text{r}}$ in (\ref{EQU:MPHAT}), which is not used in SRP-P.

As SRP-P is solely model-based (not data-driven) and the individual ADNN input is from a single microphone \cite{ZWang2019}, the ADNN of SRP-MP can be trained once, and then it can be used for any (static or dynamic) microphone architecture by adjusting the SRP-P model. Note that $M_{\text{r}}$ can nevertheless be array dependent, and unmatched array architectures during training and testing could lead to degraded results compared to the matched scenario. In contrast, as DDNNs are connected to a specific array architecture via training (e.g., fixed inter-microphone distance, number of microphones, etc.), an application to a different array cannot yield reasonable results. Consequently, retraining is required, and sometimes even modifications to the DDNN architecture are necessary due to a different number of microphones. For implementation details of SRP-MP/SRP-P,  we refer to \cite{SRP, Pyroomacoustics}.

We like to note that the computational complexity of SRP-MP is much lower than of the \ac{DNN}s. To assess computational complexity,  we follow the approach in \cite{flops} and count the number of multiplications/divisions, additions/subtractions,  i.e.,  the number of \ac{flops}.  Non-linearities are not taken into account.  A more accurate complexity analysis would require knowledge about the employed hardware.  Assuming complex numbers to be represented by a real and imaginary part,  SRP-P requires $5\cdot K\cdot Q $ real-valued divisions per time-frame to compute and apply the PHAT weighting to each microphone.  Efficiently implemented,  SRP-P requires $6 \cdot K \cdot \frac{(Q-1)^2}{2}$  operations to compute the upper-triangle components of (\ref{Equ:PHIYY}) and approximately $4\cdot K\cdot C \cdot \frac{(Q-1)^2}{2}$  operations to compute the components in (\ref{Equ:SRP}) per time-frame.  In total this sums up to approximately $ \frac{(Q-1)^2}{2}\cdot (4\cdot K\cdot C+6\cdot K) + 5\cdot K \cdot Q $ \ac{flops}.  For $K=257$, $C=37$ and $Q=4$ the number of \ac{flops} is lower than  $ 0.2 \cdot 10^6 $.  For comparison, see the \ac{flops} of the \ac{DNN}s in Table~\ref{TAB:ARCHOVERVIEW}.

\section{Proposed Frequency-Selective \ac{DOA} Estimation}
\label{CHAP:3}
In this section, we propose different systems and training strategies for signal-aware \ac{DOA} estimation using attention. A system, thereby, consists of a module for attention and another module for \ac{DOA} estimation. We compare these systems to \cite{Mack2020SigAware},
\cite{ZWang2019}, and the R-PM* concept of \cite{zhang2019robust} in the performance evaluation.
\subsection{\ac{DOA} Modules}
\label{CHAP:3.1}
In the following, we describe the proposed extensions and modifications of our preliminary work \cite{Mack2020SigAware}. In Table~\ref{TAB:ARCHOVERVIEW}, we present various variants of different DDNNs designed to investigate specific research questions. We modified the architecture from \cite{Mack2020SigAware} by including batch normalization layers \cite{ioffe2015batch} between the \ac{CNN}s as it is known to reduce training time and makes the model less prone to the initialized weights. These models are marked with a superscript B in Table~\ref{TAB:ARCHOVERVIEW}.
\begin{table}
  \centering
  \caption{Overview of all DDNNs: ``Feature'' specifies the input of the DDNN. ``BN'' specifies whether additional 2D batch normalization layers (size 64 = number of \ac{CNN} filters per layer) are between the \ac{CNN} layers. ``Size \ac{FFNN}'' specifies the sizes of the 3 feed-forward layers in the format: input-l1;output-l1;output-l2;output-l3. For the narrowband models, the size of the three feed-forward layers is given for each frequency band. ``NB'' specifies whether there is an \ac{FFNN} per frequency band (narrowband) or a single \ac{FFNN} processing all frequency bands together to map $F$ to the \ac{DOA}. ``Abbreviation'' specifies the abbreviation we use for the respective \ac{DNN}. The \ac{CNN}s have 64 filters of shape $(2, 1)$, each, a stride of 1, no padding as in \cite{Chakrabarty2017b}. We use a ReLU activation after each hidden layer and a sigmoid activation after the output layer. Between each feed-forward layer, we use a dropout of 0.5 as in \cite{chakrabarty2019multi}. The number of \ac{flops} gives the approximate number of multiplications and additions required by each of the models per time-frame (without the masking network) for $Q=4$, $K=257$,  $C=37$.}
  \label{TAB:ARCHOVERVIEW}
  \begin{tabular}{clcccccc}
  \toprule
  \# &Abbreviation & Feature  & BN & \# \ac{CNN}s &  Size \ac{FFNN} & NB   & \# FLOPS \\
  \midrule
 1&  $\text{CNN}^{\tiny{B}}_{\tiny{\Delta P}}  $ &$\Delta P$  & \cmark & $Q-2$ &   64$\cdot$257;512;512;37 & \xmark &$\approx 22\cdot 10^6$ \\
  2& $\text{CNN}^{\tiny{N,B}}_{\tiny{\Delta P}}$  &$\Delta P$  & \cmark & $Q-2$ &   64;74;74;37& \cmark  &$\approx 11\cdot 10^6$\\
   3 & $\text{CNN}_{\tiny{\Delta P}}$ & $\Delta P$  & \xmark & $Q-2$ &   64$\cdot$257;512;512;37 & \xmark    & $\approx 22\cdot 10^6$\\

4&    $\text{CNN}^{\tiny{N}}_{\tiny{\Delta P}}$  &$\Delta P$  & \xmark & $Q-2$ &  64;74;74;37 & \cmark   &$\approx 11\cdot 10^6$ \\

 5&$\text{CNN}^{\tiny{B}}_{\tiny{P}} $& $P$  & \cmark & $Q-1$ &  64$\cdot$257;512;512;37  & \xmark & $\approx 30\cdot 10^6$\\
  6&$\text{CNN}^{\tiny{N,B}}_{\tiny{P}}$  &$P$  & \cmark & $Q-1$ &   64;74;74;37 & \cmark  & $\approx 19\cdot 10^6$\\

 7& $\text{CNN}_{\tiny{P}}$& $P$  & \xmark & $Q-1$ & 64$\cdot$257;512;512;37 & \xmark    &$\approx 30\cdot 10^6$ \\
 8&$\text{CNN}^{\tiny{N}}_{\tiny{P}}$  & $P$  & \xmark & $Q-1$ & 64;74;74;37 & \cmark&  $\approx 19 \cdot 10^6$\\

   9 &$\text{FFNN}_{\tiny{\Delta P}}$ &$\Delta P$  & \xmark & 0 &   ($Q-1$)$\cdot$257;512;512;37  & \xmark    & $\approx 1.4\cdot10^6$\\
 10& $\text{FFNN}^{\tiny{N}}_{\tiny{\Delta P}}$  & $\Delta P$  & \xmark & 0&  $Q-1$;74;74;37 & \cmark  &  $\approx 4.3\cdot10^6$ \\

  11& $\text{FFNN}_{\tiny{P}}$ &$P$  & \xmark & 0&   $Q\cdot$257;512;512;37 & \xmark  &  $\approx 1.6\cdot10^6$\\

12&$\text{FFNN}^{\tiny{N}}_{\tiny{P}}$ &$P$ & \xmark & 0 &  $Q$;74;74;37 & \cmark & $\approx 4.4\cdot10^6$ \\

\bottomrule
  \end{tabular}
\end{table}
\subsubsection{Phase Vs.  Phase Difference Input} The \ac{DOA} information is, according to physical models, given in the inter-microphone phase-differences denoted as $\Delta P \in [-\pi,\pi]^{Q-1\times K\times N}$, where $\Delta P[q,k,n]=\text{MOD}(P[q+1,k,n]-P[q,k,n],\pi)$, where MOD is the modulo operator. The information of the phase differences is in the phases, however, with an additional random offset. Consequently, mapping microphone phases to the \ac{DOA} is a many-to-one mapping due to a random phase offset on the microphone phase-differences. When using the phase difference,  a frequency-dependent one-to-one mapping exists (in the absence of noise and other distortions) from the input features to the DOA.  In Table~\ref{TAB:ARCHOVERVIEW}, these \ac{DNN}s are marked with a $\Delta P$ in the ``Feature'' column. Note that using $\Delta P$ as input reduces the number of \ac{CNN} layers to $Q-2$.
\subsubsection{Parameter Reduction} The number of parameters of the DDNN \cite{Mack2020SigAware} is dominated by the weight matrix of the first feed-forward layer. The size of this layer is very large due to the multiplication of the number of \ac{CNN} filters with the number of frequency bands (see Table~\ref{TAB:ARCHOVERVIEW}). If the \ac{CNN} is removed, the number of frequency bands is only multiplied with the number of microphones resulting in a parameter reduction of approximately $\frac{Q}{64}$. Consequently, we propose a DDNN without \ac{CNN}s to investigate whether this massive parameter reduction leads to performance degradation. In Table~\ref{TAB:ARCHOVERVIEW}, these \ac{DNN}s are abbreviated with ``FFNN''.
\subsubsection{Narrowband DDNNs}
\label{CHAP:3Narrowband}
\begin{figure}
\centering
\includegraphics{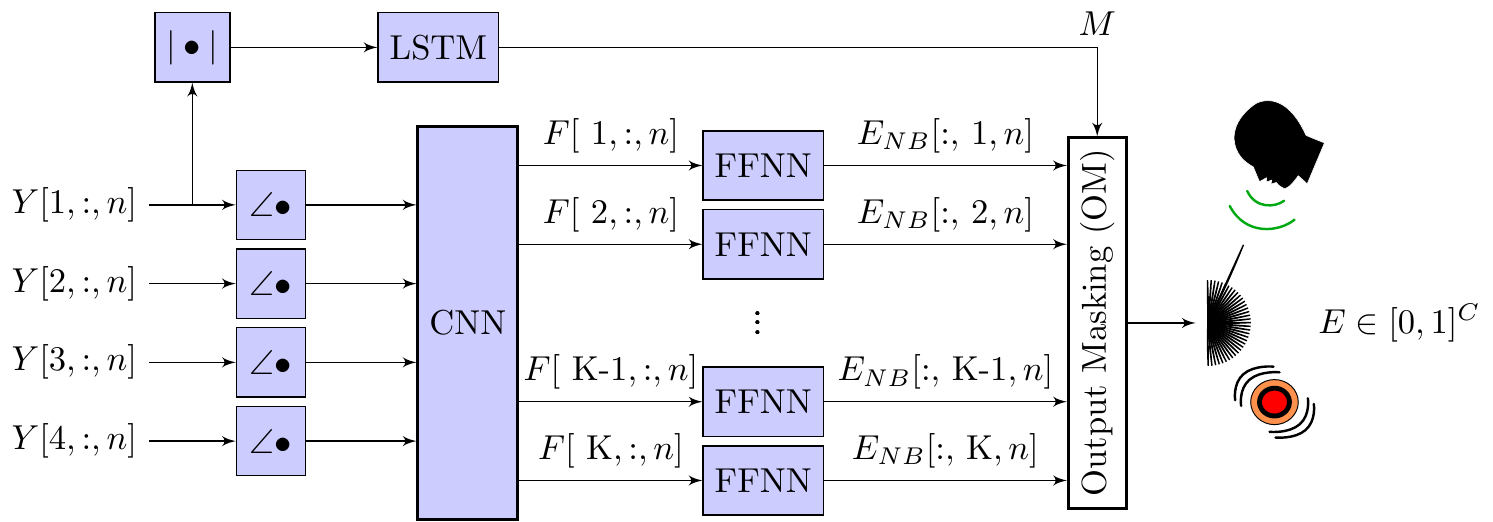}
\caption{Narrowband DDNN with an independent \ac{FFNN} per frequency band.}
\label{FIG:NARROWBAND}
\end{figure}
Many signal processing-based \ac{DOA} estimators  perform narrowband \ac{DOA} estimation and subsequently merge the narrowband estimates (e.g., via averaging \cite{NORMMUSIC}) to obtain a broadband estimate. Motivated by these approaches, we propose to use the same rationale for \ac{DOA} estimation with a \ac{DNN}.  We assume such a procedure has an additional advantage in the case of signal-aware \ac{DOA} estimation, where some frequency bands have to be disregarded. In \cite{Mack2020SigAware}, this was achieved using binary masking (B-PM, B-FM). The band selection, thereby, strongly depends on the spectral characteristics of the interference and the desired signals. Both B-PM and B-FM may introduce different kinds of noise, dependent on the spectral characteristics of the sources, in the \ac{DOA} estimation as the zeroed/randomized features are fed in the DDNN. Suppose the \ac{DOA} is estimated independently per frequency band. In that case, the \ac{DOA} estimates of the individual bands can subsequently be combined (e.g., by averaging the \ac{DOA} estimates over the frequency) to obtain a broadband estimate. Additionally, we hypothesize that such a process is more robust w.r.t. masking than B-FM and B-PM as zeroed/randomized inputs are not fed in the DDNN. As the \ac{CNN} filters only combine inter-microphone but not inter-frequency information, it is sufficient to modify the DDNN in \cite{Mack2020SigAware} such that there is an individual \ac{FFNN} with sigmoid output activation per frequency band. A scheme of this architecture is given in Figure~\ref{FIG:NARROWBAND}. In Table~\ref{TAB:ARCHOVERVIEW}, these \ac{DNN}s are marked with a \cmark in the ``NB'' column.
\subsection{Attention Module and Training Strategies}
\label{CHAP:3Attention}
For comparability, we use the same ADNN architecture for all \ac{DOA} modules (proposed and baselines). The ADNN consists of 2 long short-term memory layers (LSTM) (input dim. = 257, hidden dim. = 512) followed by a feed-forward layer with an output shape of 257 with sigmoid activation.  Per time-frame,  this model requires approximately $7.6$~million \ac{flops}\footnote{\ac{flops} of feed-forward layers are computed by doubling the multiplication of the input dimension with the output dimension to account for multiplications and additions.  An LSTM layer contains 4 feed-forward matrices of shape (input dim.$\times$ hidden dim.),  and 4 feed-forward matrices of shape (hidden dim.$\times$ hidden dim.).  \ac{flops} of CNN layers are computed by doubling the multiplication of the output dimension with the filter dimension,  and the number of filters.  Non-linearities or batch-normalization layers are not taken into account.}.  The input/output shapes are selected such that they fit the number of frequency bins per \ac{STFT} time-frame of the input.  In contrast to \cite{Mack2020SigAware}, we use an LSTM with fewer parameters instead of a BLSTM to enable online \ac{DOA} estimation. We used a dropout of 0.4 between the LSTM layers and of 0.7 before the output layer during training to avoid overfitting \cite{Srivastava2014}.
In \cite{Mack2020SigAware}, we optimized the ADNN using a speech enhancement objective. In particular, we selected frequency bands in a binary fashion to estimate the \ac{DOA}. This binary selection breaks the gradient path, as the rounding to either 0 or 1 is non-differentiable. Consequently, the ADNN cannot be trained \ac{E2E} with the \ac{DOA} estimation objective using supervised learning. Additionally, to train the ADNN \ac{E2E}, training data cannot be simulated using noise as in \cite{Chakrabarty2017a, Chakrabarty2017b} as the ADNN has to learn the spectral and temporal characteristics of the \ac{SOI}. Consequently, training data must be simulated with the \ac{SOI}. Using the time-frame-based training method in \cite{Chakrabarty2017a, Chakrabarty2017b} would require an activity detection for the \ac{SOI}. To avoid \ac{SOI} activity detection and train the ADNN and the DDNN \ac{E2E}, we propose different training strategies using ratio instead of binary attention techniques for frequency bin selection.
\subsubsection{End-to-End Training With Feature or Phase Masking}
\label{CHAP:3E2EFMPM}
\label{CHAP:E2ETFMPM}
The ADNNs yield the (single-channel) ratio attention $M_{\text{r}} \in [0,1]^{K\times N}$ for $N$ successive time-frames. We assume the source to be static, i.e., it does not move for $N$ time-frames. We propose to apply $M_{\text{r}}$ [rather than $M_{\text{b}}$ as in (\ref{EQU:B-PM}) and (\ref{EQU:B-FM})] to the features, i.e.,
\begin{equation}
  \label{EQU:R-FM}
  \widetilde{F}[k,:,n] = F[k,:,n] \cdot M_{\text{r}}[k,n],
\end{equation}
or to the input, i.e.,
\begin{equation}
    \label{EQU:R-PM}
  \widetilde{P}[q,k,n] = M_{\text{r}}[k,n] \cdot P[q,k,n]  + (1-M_{\text{r}}[k,n])\cdot \nu [q,k,n].
\end{equation}
We refer to the application of ratio attention in (\ref{EQU:R-FM}) and (\ref{EQU:R-PM}) as ratio feature-masking (R-FM) and ratio phase-masking (R-PM), respectively.
The respective DDNN output is marked as $E_{\text{FM}}$, or $E_{\text{PM}}$.  For evaluation over multiple time-frames it is common \cite{chakrabarty2019multi} to gather \ac{DOA} information across consecutive time-frames by averaging, i.e.,
\begin{equation}
  \bar{E}_\bullet[e] = \frac{1}{N}\sum_{n=1}^{N}E_\bullet[:,n] \in  [0,1]^{37},
  \label{EQU:E2ELOSSFM}
\end{equation}
where $\bullet$ specifies either FM or PM.  We propose to use the same averaging concept for training. Using several time-frames for training enables training without the necessity for a source activity detector (e.g., see \cite{hammer2020fcn}) as the \ac{DOA}s of silent time-frames average out and allow the LSTM of the ADNN to exploit temporal context for attention estimation.  Note that the  \ac{DOA} label is only based on the \ac{DOA} of the \ac{SOI} but not on the other interfering sources. Please note, in contrast to \cite{Mack2020SigAware}, $M_{\text{r}}$ provides a soft rather than binary attention. We investigate training the ADNN and the DDNN together using the \ac{CCE} loss for a \ac{DOA} objective.  In the performance evaluation, we investigate using a pre-trained DDNN with noise \cite{Chakrabarty2017a}. Subsequently, the DDNN weights are frozen, and the ADNN is trained \ac{E2E} with the DDNN (frozen weights) for a \ac{DOA} estimation objective via the \ac{CCE} loss. With these experiments we investigate whether the DDNNs are biased via training towards specific spectral source characteristics and attention distributions. 

In parallel to the present work,  \cite{subramanian2020directional} proposed to use feature masking trained end-to-end with an automatic speech recognition loss.  In particular, the authors proposed to average the \ac{CNN}  features before the \ac{FFNN} such that the \ac{FFNN} learns to map denoised features to the \ac{DOA}. To avoid bias of the \ac{FFNN}, we propose to  average the frame-wise \ac{DOA} estimates after the \ac{FFNN} for training purposes only. In that way, the DDNN still operates on a time-frame basis.  Additionally, in \cite{subramanian2020directional}, the authors estimate attention from the spatial features obtained by the CNN of the STFT phases (i.e.,  without using the magnitude information); In contrast, we estimate attention from the magnitude STFT to differentiate between the SOI and other sound sources.

\subsubsection{End-to-End Training for Narrowband Estimators}
\label{CHAP:E2ETRAININGNB}
For the narrowband DDNNs, R-PM or R-FM is not necessary, as we obtain a separate \ac{DOA} per frequency band, denoted as $E_{\text{NB}}\in [0,1]^{37\times K\times N}$. Consequently, we propose to weight the individual estimates to obtain a single broadband estimate, i.e.,
\begin{equation}
    \bar{E}_{\text{NB}}[c] = \frac{\sum_{k=1}^{K}\sum_{n=1}^{N}M_{\text{r}}[k,n]\cdot E_{\text{NB}}[c,k,n]}{\sum_{k=1}^{K}\sum_{n=1}^{N}M_{\text{r}}[k,n]}.
    \label{EQU:E2ELOSSNARROW}
\end{equation}
We refer to this weighting using a ratio mask as output masking ($\text{OM}_r$). We use the \ac{CCE} loss between the label and $\bar{E}_{\text{NB}}$ for training.
\subsubsection{DOA-Based Training using SRP-MP}
Finally, we compare the DDNN based approaches with SRP-MP. To train the ADNN  using  a \ac{DOA} objective, we propose to minimize the MSE between the SRP-P \ac{SPS} obtained from the clean non-reverberant signals $X^{\textrm{d}}_1$ (denoted as $\bar{E}^{X_1^{\textrm{d}}}_{\textrm{SRP-P}}$) and the SRP-MP \ac{SPS} obtained from the mixture signals $Y$, i.e.,
\begin{equation}
J_{\textrm{SRP-MP}} =\frac{1}{C}\sum_{c}\left( \bar{E}_{\textrm{SRP-MP}}[c]-\bar{E}^{X_1^{\textrm{d}}}_{\textrm{SRP-P}}[c] \right)^2.
\label{EQU:LOSSE2ESRP}
\end{equation}
We implemented the SRP-P code \cite{Pyroomacoustics} in PyTorch to enable training and included the attention mechanism. Note that a \ac{CCE} loss cannot be applied here, as even the SRP-P estimate of the non-reverberant, noise, and interference-free signals exhibit broad lobes with non-zero entries aside from the desired \ac{DOA}. Consequently, these non-zero entries cannot be removed in the proposed framework such that the \ac{CCE} loss cannot be applied.
\section{Data Sets}
\label{CHAP:4}
\begin{table}[t]
  \centering
%\small
%\setlength\tabcolsep{0pt} % make LaTeX figure out intercolumn spacing
\caption{Parameters of the RIRs for the different data sets.}
\begin{tabular}{c cccc}
\toprule\
&  Training &  Validation & Test Measured \cite{Hadad2014}   & Test Simulated  \\
     \midrule
$T_{60} [s]$ &\{$.2,.3,.4,.6,.8$\}&\{$.45,.6,.75$\} &  \{$.16,.36,.61$\}&  \{$.38,.7$\} \\
SMD [m] & $\{1, 2\}$ & $\{1.2, 2.3\}$ &$\{1, 2\}$& $\{1.3, 1.7\}$ \\
$\theta\:[^\circ]$ & $\{0, 5, \ldots, 180\}$ &$\{0, 5, \ldots, 180\}$ & $\{0, 15, \ldots, 180\}$ & $\{0, \frac{180}{179}, \ldots, 180\}$ \\
\bottomrule
\end{tabular}
\label{TAB:RIRCONFIG}
\end{table}

The data sets used for training, validation, and test are introduced in this section. The \ac{STFT} parameters were a sampling frequency of $16$~kHz, a hop-size of $16$~ms, and a window-length of $32$~ms such that $K=257$. Each file contains speech at its center.  Parts ($N=100$) of longer speech files were extracted based on local energy accumulations around the center to exclude silent files/files with little speech activity.  Subsequently, the files are convolved with RIRs and cut to $N=100$ ($\approx1.6$~s). 
\subsection{Room Impulse Responses}
For training, validation, and test, different RIRs were simulated using the image-method \cite{Allen1979, RIRGenerator}.  For test, also measured RIRs were used from \cite{Hadad2014}.  All RIRs specify a ULA with four microphones and an 8~cm inter-microphone distance. From \cite{Hadad2014}, we used the central four microphones from the eight microphones 8~cm configuration. The RIR parameters, like the source microphone-center distance (SMD) or the reverberation time $T_{60}$, are summarized in Table~\ref{TAB:RIRCONFIG}. The training, validation, and test RIRs were simulated in different rooms $[m]$, $\{[6,6,2.7], [5,4,2.7], [10,6,2.7], [8,3,2.7], [8,5,2.7]\}$ for training, and $\{[9,11,2.7], [10,10,2.7], [9,5,2.7] \}$ for validation, and $\{[9,4,3], [5,7,3] \}$ for the simulated test set ({\scshape{testSIMRIR}}). Note that although the SMD in training and the measured test ({\scshape{testMEASRIR}}) is the same, the different reverberation times and rooms change the direct-to-reverberation ratio such that it can be seen as unmatched conditions. {\scshape{testMEASRIR}} consists of $T_{60}\cdot \text{SMD}\cdot \text{\ac{DOA}s} = 3\cdot 2 \cdot 13=78$ RIR configurations [RIRs to all microphones].  Combining the rooms with the parameters from Table~\ref{TAB:RIRCONFIG} results in $3\cdot 50\cdot37$,  $18\cdot37$, $8\cdot180$ RIR configurations for the training, validation, and simulated test sets, where $37$ and $180$ are the numbers of \ac{DOA}s, respectively. For each simulation, the microphone array was placed randomly in the room. The source was placed according to the respective ground-truth \ac{DOA} and SMD. The constellation was rotated randomly around a random axis. The minimum distance from all microphones and the source to the wall was set to 1~m. For the training RIRs, the generation process was repeated three times.
\subsection{Single-Source Data Sets}
To investigate the influence of attention on the performance of DDNNs, we generate data sets with a single speech source and spatiotemporally white microphone self-noise with a signal-to-noise-ratio (SNR) $\in [10,30]$~dB. Speech files from the test set from Librispeech \cite{Libri2015} were convolved with the RIRs from {\scshape{testMEASRIR}}. Per microphone configuration, the process was repeated 20 times with a random SNR. For the training and validation set, we generated data by convolving the training and validation RIR sets with white noise as in \cite{Chakrabarty2017a}. From each resulting file, we selected the first $1.6~s$, which results in 100 \ac{STFT} frames per file ($\approx~1.6~s$). The total number of training, validation, and test time-frames is $11.1\cdot1e6$, $1.3\cdot1e6$, and $156\cdot1e3$ \ac{STFT} frames for the respective sets. We refer to these sets as {\scshape{train-1S}}, {\scshape{val-1S}}, and {\scshape{test-1S}}.
\subsection{Two-Sources Data Sets}
To investigate the effect of \ac{E2E} training for signal-aware \ac{DOA} estimation of the ADNN and the DDNN, we generate training, validation, and test sets consisting of two sources (signal-to-interference ratio (SIR) $\in [-6,6]$~dB) and spatiotemporally white microphone self-noise with an SNR $\in [20,30]$~dB. The first source is always a speech source from the respective set of Librispeech \cite{Libri2015}.   The second source is random interference, e.g., guitar, engine, piano, ..., from the respective sets of the YouTube-based FSDnoisy18k \cite{Gemmeke2017,fonseca2019learning}. Due to the temporal sparsity of some files in FSDnoisy18k, we computed local energy accumulations to select energy-rich source segments such that very sparse files can be excluded. Please note, the second source does not contain speech. The \ac{DOA} of the speech source is selected deterministically to yield a uniform distribution over all $C$ possible \ac{DOA}s. The \ac{DOA} of the interfering source was selected randomly from all $C$ \ac{DOA}s with a spatial separation of both sources larger than $5^{\circ}$ for evaluation purposes. The rest of the procedure is similar to the single-source case resulting in the same number of time-frames in the respective sets. We refer to these sets as {\scshape{train-2S}}, {\scshape{val-2S}}, and {\scshape{test-2S}}.  Additionally, we generated a third high-resolution test set referred to as {\scshape{test-2S-HR}} with the RIRs from {\scshape{testSIMRIR}} with a total of $\approx 2.88\cdot1e6$ \ac{STFT} frames.  Note that for this test set, the minimum angular distance between both sources is larger than $\frac{180}{179}^{\circ}$.

For training the ADNNs, we used the signals of the first microphone of the training sets.
\section{Performance Evaluation}
\label{CHAP:5}
\begin{table*}
  \renewcommand{\arraystretch}{1.1} % Default value: 1
  \caption{Evaluation of all \ac{DNN}s on {\scshape{test-1S}} for the central 0.8~s (50 \ac{STFT} frames). The first column represents a broadband evaluation. In the second and third columns, in each file, 50  randomly selected frequency bands were used (rB-PM, rB-FM) - meaning that the respective mask $M_{\text{b}}$ consists of ones in the 50 randomly selected bands and zeros, elsewhere. In the fourth column, the frequency bands 100 to 150 were used (dB-FM/dB-OM). Models~13 and 14 are DDNNs trained for \ac{DOA} estimation only (not signal-aware).  Model~13 is the DDNN basis we extended in \cite{Mack2020SigAware} for signal-aware \ac{DOA} estimation. The trained weights of the Models~13 and 14 are available on \textit{https://github.com/Soumitro-Chakrabarty/Single-speaker-localization}.  Model~15 and 16 are standard implementations of MUSIC and SRP-P from the pyroomacoustics library \cite{Pyroomacoustics} that were evaluated on the randomly/deterministically selected frequency bands. In MUSIC, we additionally used the band-wise normalization proposed in \cite{NORMMUSIC}. ``No Masking''  is equivalent to a mask that consists of ones, only. The results in the table demonstrate the effect of attention on \ac{DOA} estimators. }
\setlength\tabcolsep{4pt} % make LaTeX figure out intercolumn spacing
  \centering
\begin{tabular}{llrrrrrrrrrrrr}
\toprule
\#&Model&\multicolumn{3}{c@{}}{No Masking}&\multicolumn{3}{c@{}}{rB-PM }&\multicolumn{3}{c@{}}{rB-FM/rB-OM  } &\multicolumn{3}{c@{}}{dB-FM/dB-OM } \\ \cmidrule(lr{.75em}){3-5} \cmidrule(lr{.75em}){6-8} \cmidrule(lr{.75em}){9-11}\cmidrule(lr{.75em}){12-14}
&&   MAE &  ACC &  psACC&   MAE &  ACC &  psACC&   MAE &  ACC &  psACC &   MAE &  ACC &  psACC  \\
\midrule

1&$\text{CNN}^{\tiny{B}}_{\tiny{\Delta P}}$  & 2.2 & 70 &   \textbf{89} &  7.1 & 50 &   79 &  3.9 & 58 &   85  & 15.2 & 41 &   66 \\
2&$\text{CNN}^{\tiny{N,B}}_{\tiny{\Delta P}}$ & 2.4 & \textbf{77}  &  86 &  --- & --- &  --- &  7.4 & 57 &   79 & 52.4 & 28 &   35 \\
3&$\text{CNN}_{\tiny{\Delta P}}$ & 2.3 & 73 &   88  &  7.3 & 44 &   77 &  3.2 & 64 &   84  & 14.2 & 48 &   73 \\

4&$\text{CNN}^{\tiny{N}}_{\tiny{\Delta P}}$ & 4.5 & 67 &   87 & --- & --- &  ---  & 10.1 & 50 &   75  & 57.1 & 20 &   33 \\
\midrule
5&$\text{CNN}^{\tiny{B}}_{\tiny{P}}$  & \textbf{1.8} & \textbf{77} &   88  &  6.7 & 63 &   84  &  2.5 & 71 &   \textbf{90}  & 20.1 & 50 &   67 \\

6&$\text{CNN}^{\tiny{N,B}}_{\tiny{P}}$    & 5.8 & 59 &   77 & --- & --- &  ---  & 12.5 & 43 &   68 & 51.5& 14 &   26 \\
\bottomrule
\end{tabular}
\label{TAB:RDMASK}
\end{table*}
We evaluate the ADNNS and DDNNs using the mean absolute error (MAE), accuracy (ACC), and pseudo accuracy (psACC) metrics. The MAE is the average over several files  of the \ac{AE}, which is defined as
\begin{equation}
\text{AE} =  \left|\theta_1- \widetilde{\theta}\left[ \textrm{argmax}\left\{\sum_{n=1}^{N_{\textrm{e}} }E_\bullet[n,:]\right\}\right]\right|,
\end{equation}
where $N_{\textrm{e}} $ is the considered number of time-frames per file for evaluation in the test set. We assume a result to be accurate if the AE is smaller than $ 5^\circ$ and pseudo accurate if the AE is smaller than $ 10^\circ$ such that the neighbouring classes of the DDNN output are included. We report these metrics on a frame ($N_{\textrm{e}} =1$), a 50-frame ($\approx 0.8~s$, $N_{\textrm{e}} =50$), or 100-frame ($\approx 1.6~s$, $N_{\textrm{e}} =100$) basis to show the performance of the algorithms over different context lengths.    The desired length $N_{\textrm{e}}$, thereby, depends on the application.  For tracking fast moving sources,  for example,  a small $N_{\textrm{e}}$ is required to provide sufficient temporal resolution.  For slowly moving or static sources,  a large $N_{\textrm{e}}$ might be beneficial to increase the \ac{DOA} estimation accuracy.  As the focus of the paper is the localization of static sources, we base most of our experiments on $N _{\textrm{e}}=50$ or $N _{\textrm{e}}=100$.  Unless stated differently, all \ac{DOA} estimators (signal processing and \ac{DNN}-based) sample the \ac{DOA} space with a resolution of $5 \text{ degrees}$.
\subsection{Impact of Binary Frequency Selection on \ac{DOA} Estimation}
\label{SUBSEC:PEV1}
\begin{figure*}
\centering
\input{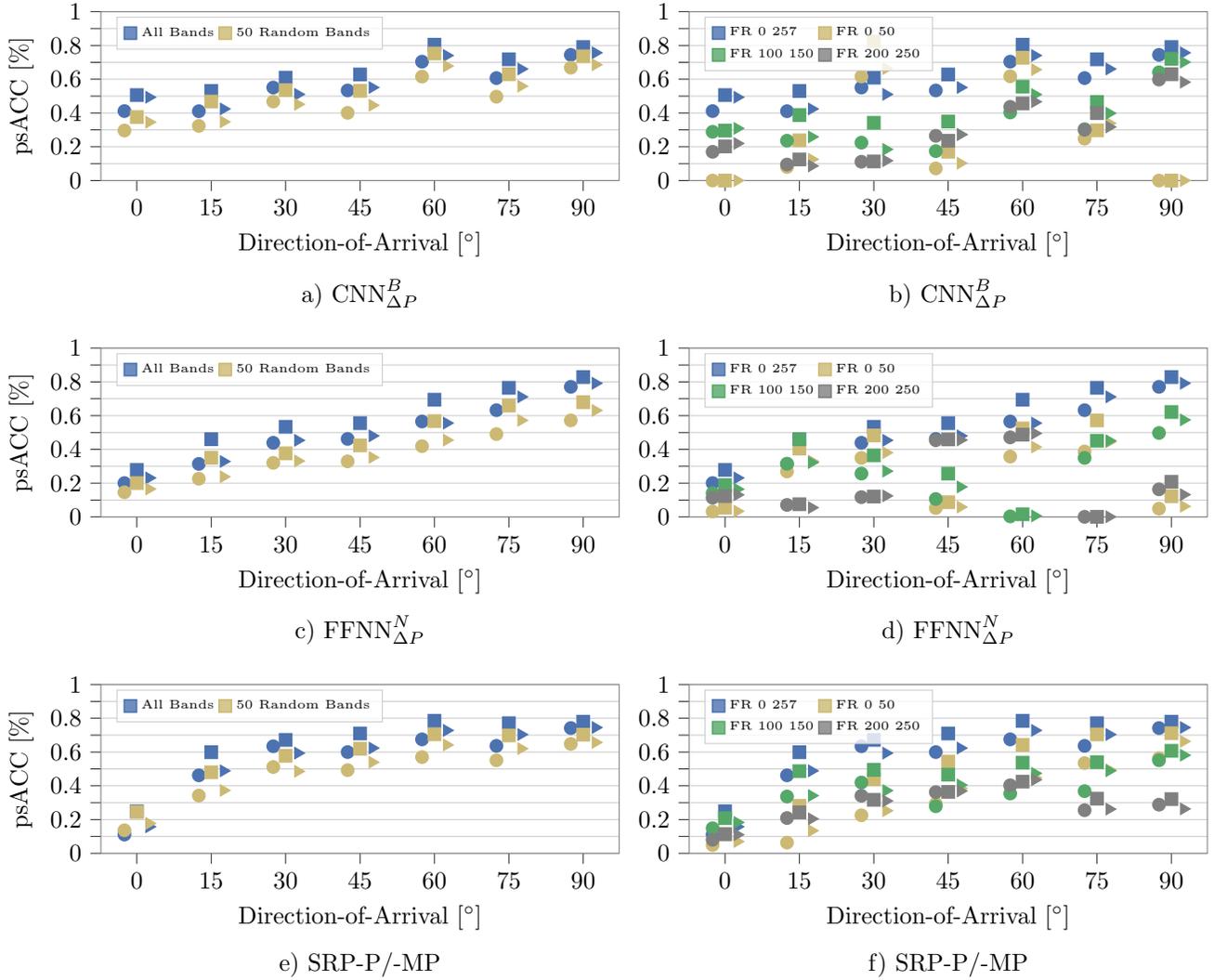}
\caption{Frame-wise psACC results over the \ac{DOA} for different reverberation times in {\scshape{test-1S}} (only the central frame where $n=50$ was used for evaluation to ensure speech activity). The square represents a $T_{60}$ of $0.16$~s, the triangle of $0.36$~s, and the circle of $0.61$~s. In the left column, B-FM/B-OM was applied such that only 50 randomly selected frequency bands were used. In the right column, B-FM/B-OM was applied for different frequency ranges (FRs), from $k=0$ to $k=50$, from $k=100$ to $k=150$, and from $k=200$ to $k=250$ were evaluated. In the respective scenarios, only frequency bands in the respective range were used. }
\label{Fig:rdFmask}
\end{figure*}

The effect of attention in terms of masking on the DDNNs has not yet been studied in depth.  In particular, it is not clear what effect different attention distributions have on the DDNN performance/whether there is a bias in the DDNN towards specific attention distributions.  To investigate the effect in a controlled way,  we evaluate two different attention distributions represented by binary masks that contain $50$ ones and $257-50=207$ zeros per time frame.  In a file with multiple time-frames, the mask is the same for all time-frames.  In the first attention distribution, the ones are selected randomly via a uniform distribution per file.  In the second attention distribution, the ones are selected deterministically.  We refer to these attention distributions as attention-distribution~one and attention-distribution~two,  respectively.  We distinguish three different binary masking procedures, (I) random phase-masking (rB-PM) and (II) random feature-masking (rB-FM) using attention-distribution~one and (III) deterministic feature-masking (dB-FM) using attention-distribution~two, where the binary mask contains ones in the frequency bands from 100 to 150.  The results are summarized in Table~\ref{TAB:RDMASK}.  The results in this section are based on the single-source data set {\scshape{test-1S}} and the DDNNs are trained with {\scshape{train-1S}}.

If no masking (i.e.,  $M=1$) is applied, all methods (except Model~6) achieve a psACC higher than or equal to $80\%$ in Table~\ref{TAB:RDMASK} in the presence of noise and reverberation. The signal-processing methods perform comparably to the deep-learning methods; see Models~2,  5,  and 15, 16. Using the inter-microphone phase-differences as input instead of the raw phases does not result in better performance for the broadband \ac{CNN}-based DDNNs. The phase difference input only improves the results for the DDNNs based solely on feed-forward layers. This shows that \ac{CNN}s can denoise the input better than \ac{FFNN}s.   In particular,  Model~5, which uses the raw phase as input, performs best in terms of ACC and MAE. A possible explanation is the additional convolution layer of the \ac{CNN}s, which use the raw phases instead of the phase differences as input that increases the learning capabilities of the DDNN. Batch normalization between the \ac{CNN} layers does not influence the results, as the performance of Models~1 and 5 is comparable to the performance of Models~3 and 7.

When masking is applied, rB-FM outperforms rB-PM consistently, as in \cite{Mack2020SigAware}. The results show that if only feed-forward layers are used, the performance drops when any type of masking is applied, as can be seen in the rB-PM column of Models~9 and 11. Consequently, we exclude these models from further evaluations. Using rB-FM, the narrowband models perform worse than the broadband models. Therefore, having inter-frequency connections in the DDNN helps to estimate the \ac{DOA}, especially given noisy inputs. Interestingly, using the phase differences instead of the phases as input improves the performance for the narrowband models.   For dB-FM, the narrowband models fail completely. Also, the performance of the broadband models is strongly degraded, although having access to the same number of frequency bins as when using rB-FM. The performance degradation is much less severe for signal processing-based methods like MUSIC or SRP-P. This effect can be explained by \ac{DNN} training. The DDNNs have been trained with spatiotemporally white noise and directional temporally white noise.  On a short time-frame basis ($\approx 32$~ms),  the addition of two white noise processes leads to time-frequency bins where the relative energy of directional and non-directional noise can vary strongly, meaning that time-frequency bins where the directional noise source is dominant are uniformly distributed over the frequencies. This distribution is resembled by rB-FM, where the bin-wise SNR for masked bins (i.e.,  unattended) can be assumed to be $-\infinity$~dB, corresponding to the time-frequency bins dominated by the non-directional noise during training.  Consequently, the bin-wise SNR distribution during testing is very different when employing dB-FM compared to the uniform bin-wise SNR distribution during training.

To investigate the effect further, we plot the psACC on a frame-basis over the \ac{DOA} and for different reverberation times in Figure~\ref{Fig:rdFmask} for selected models. For rB-FM, on the left side, the performance of the DDNNs and SRP-P deteriorates slightly. Please note the bias to $90^\circ$ is typical for a ULA. For dB-FM, shown on the right side of Figure~\ref{Fig:rdFmask}, the DDNNs fail to estimate specific directions correctly, completely for all tested reverberation times. However, the results of SRP-P do not exhibit such complete failures for specific \ac{DOA}s for different frequency distributions. The performance of the DDNNs deteriorates for all three considered frequency-band ranges in dB-FM compared to rB-FM, although having the same number of frequency bands to estimate the \ac{DOA} from. Remarkably, when using dB-FM, the performance deterioration of the DDNNs is \ac{DOA} and frequency-band range dependent. For specific \ac{DOA}s and band ranges, e.g., band range 100-150, $60^\circ$ \ac{DOA}, $\text{FFNN}_{\Delta P}^{\text{N}}$, the psACC drops to almost 0, whereas for other \ac{DOA}s, e.g., $15^\circ$, the performance is comparable to no masking. This effect is especially prominent for the narrowband model, less for the broadband model, and again less for SRP-P as it did not require any training. Consequently, having access to all frequencies as the broadband model provides more reliable \ac{DOA} estimates. Based on these findings, we exclude the narrowband estimators from further evaluations and investigate training for speech. Additionally, the performance gap between dB-FM and rB-FM suggests that training the DDNNs with the signal class to estimate the \ac{DOA} from and with the spectral attention distribution used in the test instead of uniformly distributed noise can further improve the performance. We further evaluate this hypothesis in the next section.  
\subsection{Impact of Training Data and Attention Application on  Signal-Aware \ac{DOA} Estimation}
\label{CHAP:E2EEVALUATION}
\begin{table}
  \centering
  \caption{Results for different attention schemes with different training methods for the DDNN. The ADNN is always trained \ac{E2E} with the respective attention technique (R-FM, R-PM, R-PM*) and the respective DDNN. Note that when the DDNN was trained with noise \cite{Chakrabarty2017a}, the DDNN weights were frozen for the ADNN training.}
  \begin{tabular}{ccccccc}
  \toprule
  DDNN Training &\multicolumn{3}{c}{Noise \cite{Chakrabarty2017a}}  &\multicolumn{3}{c}{E2E (\ref{EQU:E2ELOSSFM}, \ref{EQU:E2ELOSSNARROW}, or \ref{EQU:LOSSE2ESRP})}\\
  \cmidrule(lr{.75em}){2-4}  \cmidrule(lr{.75em}){5-7}  
Masking &MAE  & ACC & psACC &MAE & ACC & psACC    \\
\midrule
No Mask &37.9&42&47&16.7&63&74\\
R-FM &31.6&48&53 &\textbf{6.2}&\textbf{70}&\textbf{83}\\
$\text{R-PM*}$\cite{zhang2019robust}&---&---&---&9.0&67&80\\
R-PM &32.6&48&53&13.3&65&76\\
\bottomrule
  \end{tabular}
    \label{TAB:RESE2EDDNN}
\end{table}

\begin{table}
  \centering
  \caption{Results for different training schemes for the ADNN (MSE and DOA-based with the respective \ac{DOA} module). The DDNN was trained with speech. The SRP-P results without masking were $39.4^\circ$, $37$~\%, $43$~\%, for MAE, ACC, and psACC, respectively. For comparison,  we use our ADNN architecture for the baseline trained with PSM and use the same mask for all microphone channels instead of the channel-dependent masks as in \cite{ZWang2019} (results for microphone channel-dependent masks as in \cite{ZWang2019} are $6.9^\circ$, $67$~\%, $80$~\% for MAE, ACC and psACC, respectively).   }
  \label{TAB:RESE2EADNN}
  \begin{tabular}{cccccccccc}
  \toprule
  ADNN Training &\multicolumn{3}{c}{MSE (\ref{EQU:MSESPEECHENHANCEMENT})}  &\multicolumn{3}{c}{E2E/SPS (\ref{EQU:E2ELOSSFM}, \ref{EQU:E2ELOSSNARROW}, or \ref{EQU:LOSSE2ESRP})}&\multicolumn{3}{c}{PSM (\cite{ZWang2019})}\\
  \cmidrule(lr{.75em}){2-4}  \cmidrule(lr{.75em}){5-7}  \cmidrule(lr{.75em}){8-10}  
Method &MAE & ACC & psACC &MAE & ACC & psACC  &MAE & ACC & psACC    \\
\midrule
SRP-MP& 9.0&62& 78&6.7&69&82&6.8&67&81\\
R-FM &24.5&51&62&\textbf{6.2}&\textbf{70}&\textbf{83}&-&-&-\\
\bottomrule
  \end{tabular}
\end{table}

Here we compare different training strategies for the \ac{DOA} and attention modules in the presence of two sources (a directional speech and a directional interference source) and noise.  For fully \ac{DNN}-based \ac{DOA} estimation, we evaluate attention application in the form of the proposed R-FM and R-PM and compare to R-PM* \cite{zhang2019robust}.  Training strategies include: (I) Training the ADNN and the DDNN jointly E2E using the \ac{CCE} loss for the \ac{DOA} label of speech with {\scshape{train-2s}} (see Equation (\ref{EQU:E2ELOSSFM})).  (II) Training the DDNN on a time-frame basis with directional noise sources from {\scshape{train-1S}} \cite{Chakrabarty2017a}, freezing the DDNN weights and training the ADNN E2E using {\scshape{train-2s}} and the \ac{CCE} loss with the \ac{DOA} label of speech  (see Equation (\ref{EQU:E2ELOSSFM})).  (III) Training an ADNN  with the MSE for speech enhancement and combining it with the DDNNs from (I) and (II).  (IV) Training the DDNN without attention using {\scshape{train-2s}} and the \ac{CCE} loss with the \ac{DOA} label of speech.  (see Equation (\ref{EQU:E2ELOSSFM})).  (V) Training the DDNN without attention and directional noise sources \cite{Chakrabarty2017a}.   For hybrid systems,  we compare ADNNs combined with SRP-MP trained using the MSE, or the PSM for speech enhancement with the proposed training based on the \ac{SPS} using {\scshape{train-2s}}.  We refer to the respective methods via SRP-P (MSE), SRP-P (PSM), SRP-P (SPS), respectively. The evaluation was performed using data simulated with measured RIRs using {\scshape{test-2S}}. Note that the ground-truth label is only based on the \ac{DOA} of the \ac{SOI}. Unless specified differently, the used DDNN is $\text{CNN}^{\tiny{B}}_{\tiny{P}}$.

\begin{figure}
\centering
\input{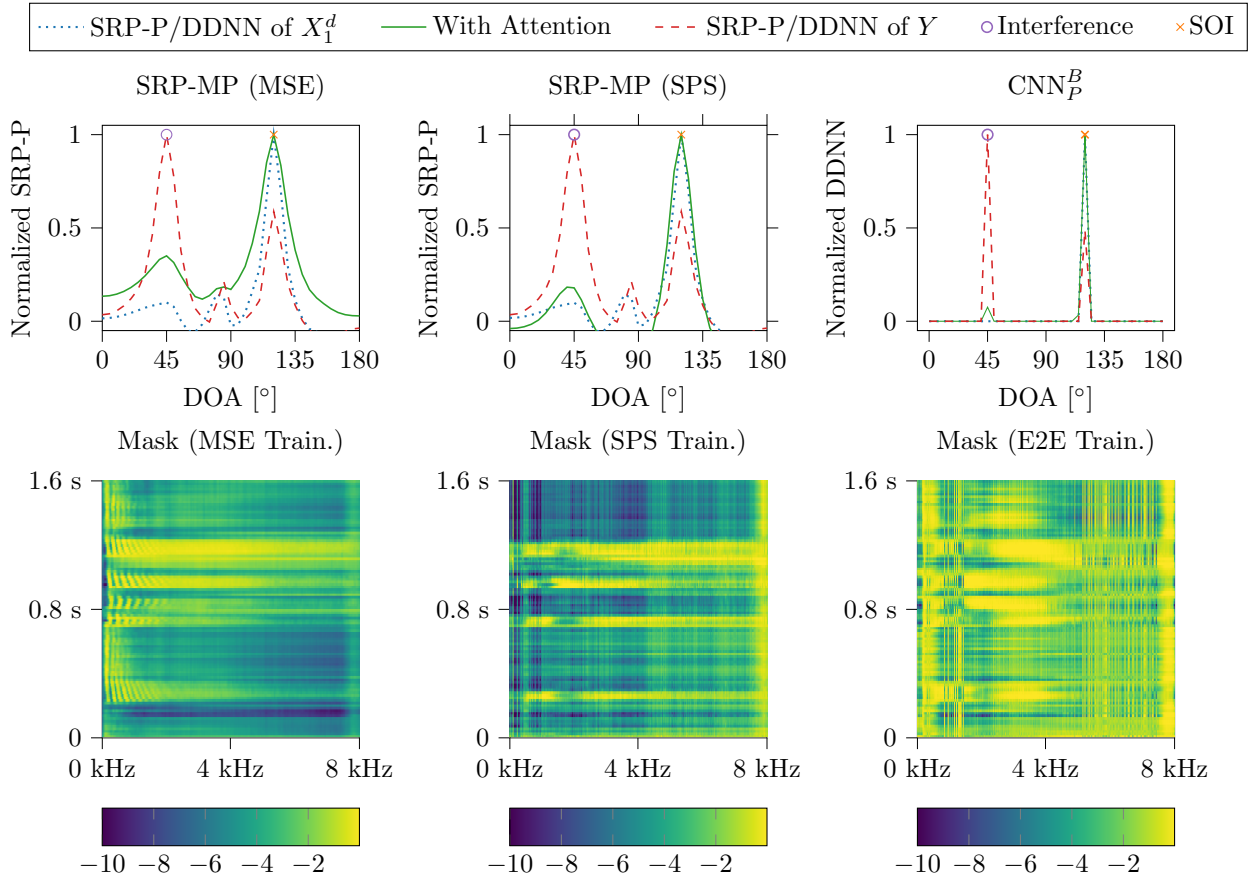}
\caption{Outputs of SRP-MP and the DDNN (R-FM)  in the first row with the respective masks (dB) in the second row.  The ADNNs are trained with the MSE (column 1),  using the \ac{SPS} with SRP-MP (colum 2) or \ac{E2E} with the DDNN (column 3). Label and Interference specify the ground-truth \ac{DOA}s of the two present sources, speech and non-speech, respectively. Note that the temporal evolvement of the mask is shown on the y-axis, to simplify a comparison of the MSE and the other masks.  }
\label{Fig:SRP}
\end{figure}

In Table~\ref{TAB:RESE2EDDNN}, we report results for the different training strategies for the DDNN.  The worst results are expected with training strategies (IV) and (V) as no attention is applied.  The DDNN trained with noise in (V) cannot decide whether to focus on the speech or the noise source but estimates the \ac{DOA}s of both sources. By chance, either the speech or the noise \ac{DOA} is picked in the evaluation (see, for example, the red line of the \ac{CNN} in Figure~\ref{Fig:SRP}).  Note that (V) only serves as an anchor that puts the results into perspective.  As it is not performing signal-aware \ac{DOA} estimation, it does not serve as a baseline.  When training with speech in (IV) instead of noise, the DDNN seems to learn speech-specific structures in the phase map such that the \ac{DOA} of the speech source is more prominent than the \ac{DOA} of the noise source.  As a result,  when training the DDNN with speech rather than noise, the MAE is reduced from $37.9^\circ$ to $16.7^\circ$ when no mask is applied. Consequently, the DDNN learned speech-specific characteristics from the input to be biased to speech sources. Note that no magnitude information is used here. This bias cannot be learned when training with spatiotemporally white Gaussian noise \cite{Chakrabarty2017a} as in (V).

When additional attention in the form of R-PM, R-PM*, or R-FM is applied, the results improve further. The best performance is achieved using the proposed R-FM. The application of R-FM yields better results than R-PM* \cite{zhang2019robust}. This shows that the masking of features after the \ac{CNN} layers is superior to masking the input. Interestingly, the DDNN achieves better results for R-PM* compared to R-PM. When the mask is zero at some bins in R-PM*, this corresponds to a \ac{DOA} of $90^\circ$ assuming the DDNN learns a physical model. The additional noise included by the unlearnable phase randomization of R-PM seems to deteriorate the results more severely than the deviation of the physical model when using R-PM*. This result and the strong discrepancy between the DDNN performance for rB-FM and dB-FM in Section~\ref{SUBSEC:PEV1} show that the DDNN does not learn an exact physical model. Instead, it learns the mapping from input to the \ac{DOA} that optimizes the loss based on the training data.

In Table~\ref{TAB:RESE2EADNN}, we report the results of different training methods for the ADNN, namely the MSE (\ref{EQU:MSESPEECHENHANCEMENT}), the PSM (\ref{EQU:PSM}) and with the respective \ac{DOA} module  using the SPS loss for SRP-MP (\ref{EQU:LOSSE2ESRP}) and \ac{E2E} for the DDNN (\ref{EQU:E2ELOSSFM}). All methods improve the result compared to the attention-free scenarios. For SRP-MP, training the ADNN with a localization loss yields slightly better results than training with the MSE.  Training with the PSM further minimizes the gap.  This can be explained as the PSM takes the phase into account, which is important for source localization.  Additionally, the MSE weights mask differences with the mixture magnitude.  This increases the weight of low frequencies if the \ac{SOI} is speech, although higher frequencies allow for more accurate localization.  For the DDNN, \ac{E2E} training is more important, as the MSE mask yields an MAE of $24.5^\circ$, and the \ac{E2E} mask an MAE of $6.2^\circ$. Note that the computational complexity of SRP-MP is much less than for the DDNN, but the performance is comparable. Also, SRP-MP does not have a bias towards the respective training source in the \ac{DOA} module in contrast to the DDNN.

Especially interesting is the small performance difference for SRP-MP when a PSM or an \ac{SPS} mask is used. In terms of dual-use, this approach allows using independently trained source separation/enhancement/extraction ADNNs with signal processing-based methods for \ac{DOA} estimation without strong performance degradation in terms of \ac{DOA} estimation compared to \ac{SPS} training or \ac{E2E} training using the DDNN. Another advantage of the SRP-MP~(PSM) approach is the independence of the training data of the array architecture, as only an ADNN and no DDNN is used. When DDNNs are used, retraining is required for every new array architecture, whereas with SRP-MP, the SRP-P model can be adjusted w.r.t. the array architecture.  Additionally, when the ADNN is trained with the PSM, it is sufficient to simulate training data for a single microphone and not for an array, which reduces the overall computational burden when SRP-MP~(PSM) is trained. Furthermore, when training the ADNN with the MSE/PSM, the signal processing-based method can be used as a black box and does not have to be implemented in a differentiable way. An advantage of the DDNN is that it only requires the \ac{DOA} label of the \ac{SOI} for training such that it could be trained with measured data if the ground-truth \ac{DOA} is provided, e.g., by an optical tracking system. In contrast, all investigated training objectives for the ADNN using SRP-MP require a representation of the \ac{SOI} at the first microphone, which complicates training using measured data.

\begin{figure}
  \centering
  \begin{subfigure}[b]{0.45\textwidth}
    \centering
% This file was created by tikzplotlib v0.9.4.
\begin{tikzpicture}[trim axis left, trim axis right]
% \useasboundingbox (0,0) rectangle (6cm,4cm);
\definecolor{color0}{rgb}{0.12156862745098,0.466666666666667,0.705882352941177}
\definecolor{color1}{rgb}{1,0.498039215686275,0.0549019607843137}
\definecolor{color2}{rgb}{0.172549019607843,0.627450980392157,0.172549019607843}
\definecolor{color3}{rgb}{0.83921568627451,0.152941176470588,0.156862745098039}
\definecolor{color4}{rgb}{0.580392156862745,0.403921568627451,0.741176470588235}

\begin{axis}[
legend cell align={left},
legend style={fill opacity=0.8, draw opacity=0.5, text opacity=1, draw=white!80!black, only marks},
tick align=outside,
tick pos=left,
xtick = {-1, 0,1,2,3,4,5,6,7,8,9},
ytick = {0, 10,  20, 30,40,50},
xticklabels = { 0,.1, .2, .3,.4,.5,.6,.7,.8,.9},
xlabel = {$v_{thr}$},
ylabel = {MAE [$^\circ$]},
grid=both,
title={},
x grid style={white!69.0196078431373!black},
xmin=-1, xmax=8,
xtick style={color=black},
y grid style={white!69.0196078431373!black},
ymin=0, ymax=50,
ytick style={color=black},
width=5cm%9.2cm
]
\addplot [dotted,thick, color1]
table {%
8 30.4
7 18.2
6 11.8
5 8.6
4 6.7
3 6.2
2 7.5
1 10.2
0 18.2
-1 39.4
};
\addlegendentry{SRP-MP}

\addplot [dashed,thick, color0]
table {%
8  84.0
7  70.0
6  55.0
5 40.0
4 28.4
3 20.5
2 14.9
1 15.2
0  19.8
-1 37.9
};
\addlegendentry{$\text{CNN}^{\tiny{B}}_{\tiny{P}}$}

\addplot [dashed,thick, black]
table {%
8  6.2
7  6.2
6  6.2
5 6.2
4 6.2
3 6.2
2 6.2
1 6.2
0  6.2
-1 6.2
};

\end{axis}
\begin{axis}[
  legend cell align={left},
  legend style={fill opacity=0.8, draw opacity=0.5, text opacity=1, draw=white!80!black, only marks},
  tick align=outside,
  tick pos=left,
  xtick = {-1, 10},
  ytick = {6.2,  14.9},
  xticklabels = {,,,,,,,,},
  grid=both,
  x grid style={white!69.0196078431373!black},
  xmin=0, xmax=8,
  xtick style={color=black},
  y grid style={white!69.0196078431373!black},
  ymin=0, ymax=50,
  ytick style={color=black},
  width=5cm,
ytick pos = right
]
\end{axis}
\end{tikzpicture}
\end{subfigure}
  \begin{subfigure}[b]{0.45\textwidth}
  \centering
% This file was created by tikzplotlib v0.9.4.
\begin{tikzpicture}[trim axis left, trim axis right],
% \useasboundingbox (0,0) rectangle (6cm,4cm);
\definecolor{color0}{rgb}{0.12156862745098,0.466666666666667,0.705882352941177}
\definecolor{color1}{rgb}{1,0.498039215686275,0.0549019607843137}
\definecolor{color2}{rgb}{0.172549019607843,0.627450980392157,0.172549019607843}
\definecolor{color3}{rgb}{0.83921568627451,0.152941176470588,0.156862745098039}
\definecolor{color4}{rgb}{0.580392156862745,0.403921568627451,0.741176470588235}

\begin{axis}[
legend cell align={left},
legend style={fill opacity=0.8, draw opacity=0.5, text opacity=1, draw=white!80!black, only marks},
tick align=outside,
tick pos=left,
xtick = {-1,0,1,2,3,4,5,6,7,8,9},
ytick = {0,10,20,30,40,50,59,70,80,90,100},
yticklabels = {0,,20,,40,,59,,80,,100},
xticklabels = {0,.1, .2, .3,.4,.5,.6,.7,.8,.9},
xlabel = {$v_{thr}$},
title={},
x grid style={white!69.0196078431373!black},
xmin=-1, xmax=8,
grid=both,
xtick style={color=black},
y grid style={white!69.0196078431373!black},
ymin=0, ymax=100,
ytick style={color=black},
width=5cm,%9.2cm
ylabel = {ACC/psACC},
]
\addplot [dotted,thick, color1]
table {%
8 16
7 31
6 42
5 51
4 57
3 60
2 61
1 61
0 55
-1 37
};
\addplot [dashed,thick, color0]
table {%
8 8
7 13
6 20
5 30
4 40
3 48
2 55
1 59
0 58
-1 42
};

\addplot [dashed,thick, color0]
table {%
8  9
7  18
6  30
5 44
4 55
3 64
2 70
1 70
0  66
-1 47
};

\addplot [dotted,thick, color1]
table {%
8  32
7  52
6  63
5 71
4 75
3 79
2 78
1 76
0  67
-1 43
};

\addplot [dashed,thick, black]
table {%
8  83
7  83
6  83
5 83
4 83
3 83
2 83
1 83
0  83
-1 83
};

\addplot [dashed,thick, black]
table {%
8  70
7  70
6  70
5 70
4 70
3 70
2 70
1 70
0  70
-1 70
};
\end{axis}

\begin{axis}[
  tick align=outside,
  xtick = {-1,10},
  tick pos = left,
  ytick = {61,79},
  xticklabels = {},
  grid=both,
  x grid style={white!69.0196078431373!black},
  xmin=0, xmax=8,
  xtick style={color=black},
  y grid style={white!69.0196078431373!black},
  ymin=0, ymax=100,
  ytick style={color=black},
  width=5cm,
ytick pos = right
]
\end{axis}

\end{tikzpicture}
\end{subfigure}
\caption{Results of MAE, ACC, and psACC of B-FM \cite{Mack2020SigAware} and SRP-MP. The ADNN is trained with the MSE (\ref{EQU:MSESPEECHENHANCEMENT}) for a speech enhancement objective and the DDNN is trained with noise \cite{Chakrabarty2017a}. Note that the psACC is always higher than the ACC. The black lines mark the results of R-FM from Table~\ref{TAB:RESE2EDDNN}.}
\label{Fig:BINARYATTENTION}
\end{figure}
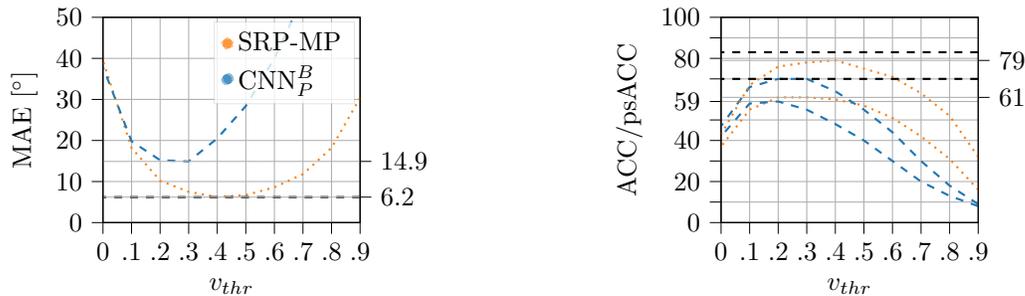
In Figure~\ref{Fig:SRP}, we show the \ac{DOA} estimation performance based on SRP-MP coefficients and based on a DDNN with the respective attention masks. Training \ac{E2E}/with the \ac{SPS} yields different masks that cannot be used for an enhancement objective. For example, the masks trained with the speech enhancement objective clearly exhibit the magnitude structure of speech at lower frequencies, whereas the masks trained for localization do not. Especially interesting are the differences between the masks obtained with the \ac{SPS} training for SRP-MP and the \ac{E2E} training with the DDNN. Where the SRP-MP~(SPS) mask is relatively sparse, the DDNN mask is not. In the DDNN, all frequency bands are connected and yield a single estimate per time frame. SRP-MP can be interpreted as estimating a \ac{DOA} per frequency bin and averaging these estimates. In the DDNN case, the internal connection seems to require rather non-sparse inputs, whereas the averaging of SRP-MP~(SPS) seems to select only a few time-frames but then nearly all of the respective frequency bins. A comparison of the masks of  SRP-MP~(SPS) and  SRP-MP~(MSE) shows that  SRP-MP~(SPS) yields masks that focus on the low-reverberant speech onsets. The SRP-MP~(MSE) mask is between the SRP-MP~(SPS) and the DDNN masks in terms of sparsity. The SRP-MP coefficients of MSE and \ac{SPS} masks look very similar, although the respective masks are quite different. This shows the robustness of SRP-MP w.r.t. different input masks. The DDNN estimates look sharper than the SRP-MP estimates. The DDNN was trained to have sharp outputs and prior information of an angular separation of sources larger than $5^\circ$ from the training data, whereas SRP-MP is model-based and not optimized to yield such sharp outputs. The different output representations, consequently, may be misleading in terms of selecting the best algorithm. This is justified as the respective ACC, psACC, and MAE results are very comparable for the DDNN and SRP-MP in Table~\ref{TAB:RESE2EADNN}.

In Figure~\ref{Fig:BINARYATTENTION}, we report the results of B-FM as in \cite{Mack2020SigAware} over $v_{\text{thr}}$ for a DDNN trained with noise with {\scshape{train-1S}} and an ADNN trained with the MSE (\ref{EQU:MSESPEECHENHANCEMENT}) with {\scshape{train-2S}}. Note that the results differ from \cite{Mack2020SigAware}, as the ADNN is a two-layer LSTM instead of a three-layer BLSTM, the file length is different ($1.6$ instead of $3~s$). We use the same binary mask with SRP-MP~(MSE) to see whether binary or ratio masks are advantageous for SRP-MP~(MSE). We compare the results to the DDNN. The DDNN achieves the best performance at $v_{\text{thr}} \approx 0.2$. In total, the ACC is improved from $42$ to $59$~\%, and the MAE is reduced from $38^\circ$ to $14.9^\circ$ for the DDNN. For SRP-MP, the best performance is achieved at $v_{\text{thr}} \approx 0.4$ when the MAE is reduced from $39^\circ$ to $6.2^\circ$. This result is the same for the best model, the DDNN, with \ac{E2E} training in Table~\ref{TAB:RESE2EDDNN}. The ACC of SRP-MP, however, is still $9$~\% percent worse compared to the \ac{E2E} trained DDNN. The psACC of both is comparable with $79$ to $83$~\%. The performance gap of DDNN and SRP-MP in Figure~\ref{Fig:BINARYATTENTION} shows that DDNNs are more sensitive to attention than SRP-MP. As signal processing-based methods for \ac{DOA} estimation are typically not tailored to any source class or array architecture, unlike DDNNs and their computational complexity is typically less than that of DDNNs, we believe that the combination ADNNs for attention and signal processing-based methods for \ac{DOA} estimation constitutes a low-complexity solution with high psACC and low MAE.
\subsection{Evaluation of Off-Grid \ac{DOA}s}
\label{CHAP:FG}
\begin{figure}
\centering
\includegraphics{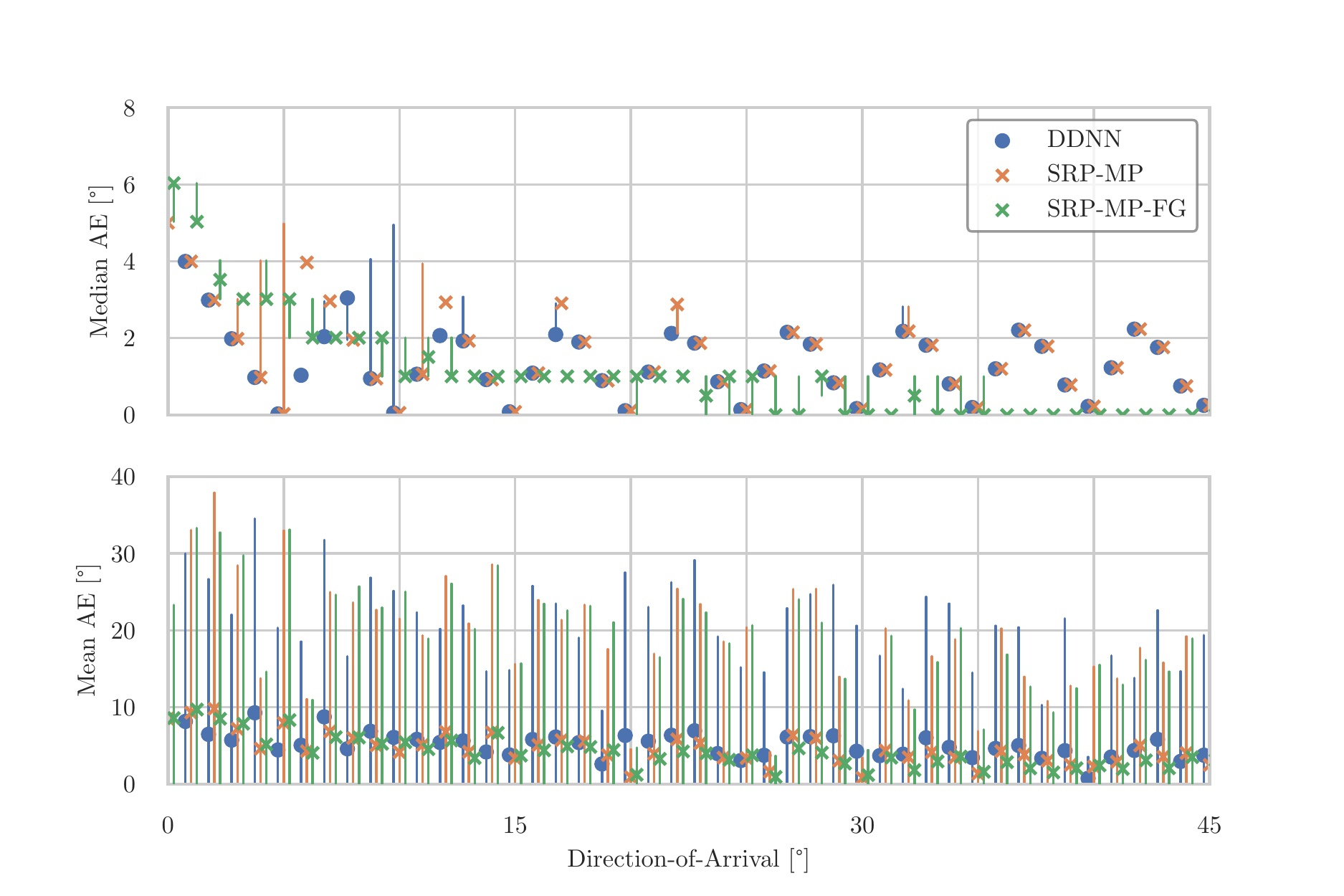}
\caption{Off-grid  median+95\% confidence interval (upper plot) and mean+standard deviation (lower plot) evaluation of the \ac{AE}. The DDNN is trained  \ac{E2E} using R-FM with a \ac{DOA} resolution of $5 ^\circ$. SRP-MP is trained using the \ac{SPS} with the resolution of $5 ^\circ$. In SRP-MP-FG, the same ADNN is used as for SRP-MP, but the resolution of the \ac{DOA} module is increased in the test to match the \ac{DOA} resolution of the simulated sources ($\frac{180}{179} ^\circ$). }
\label{FIG:OFFGRID}
\end{figure}
\begin{figure}
\centering
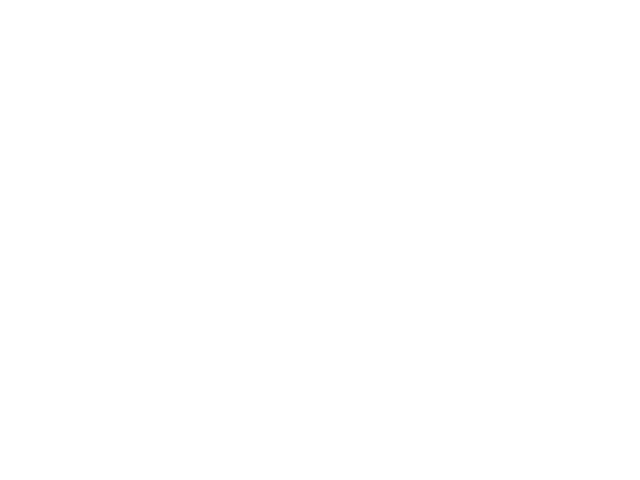
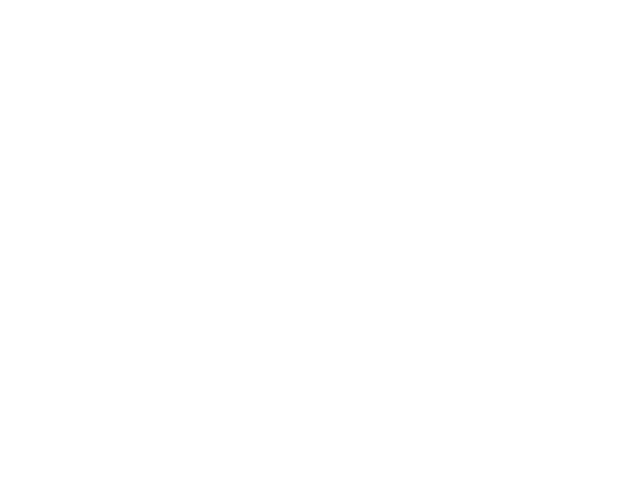
\caption{Excerpt of the confusion matrix from $0$ to $85$~degrees of the  \ac{DOA} labels vs.  the estimated  \ac{DOA}s  of the DDNN~(top) and SRP-MP~(bottom).  The estimated \ac{DOA}s have a resolution of 5~degrees and the labels of 1~degree.  The evaluated two-source data set is similar to {\scshape{test-2S-HR}}; only the label resolution was changed from  $\frac{180}{179}^\circ$  to $1^\circ$ for visualization purposes.}
\label{FIG:CONFMAT}
\end{figure}

In the previous experiments using {\scshape{test-2S}}, the \ac{SOI} \ac{DOA}s were matched to the directions that the \ac{DOA} modules sample. As all the evaluated \ac{DOA} estimation algorithms sample the \ac{DOA} space, such an on-grid comparison is fair in the sense that no method has an advantage over the other.  Also, the use of measured RIRs for data simulation inherently leads to slight off-grid positions, e.g.,  due to measurement or microphone placement errors.  However, stronger off-grid \ac{DOA}s of the sources-of-interest lead to additional rounding errors and consequently to an increased MAE. Additionally, as DDNNs learn an input to output mapping, it is not inherently clear whether they generalize to off-grid \ac{DOA}s.  

In this section,  we show that the proposed methods generalize to off-grid \ac{DOA}s.  As the baseline in \cite{ZWang2019} uses a signal processing method for \ac{DOA} estimation and the baseline in \cite{zhang2019robust} uses the same \ac{DOA} \ac{DNN} architecture as  $\text{CNN}^{\tiny{B}}_{\tiny{P}}$,  the results for off-grid \ac{DOA}s can be transferred to the respective methods and are not shown here.  We evaluate the effect of off-grid \ac{DOA}s on the signal-aware \ac{DOA} estimation performance using the two-source data set {\scshape{test-2S-HR}}. The ADNNs are trained \ac{E2E}/using the SPS with the respective \ac{DOA} module using {\scshape{train-2S}}. We show the results of the MAE and the median absolute-error (MedAE) over the \ac{DOA} in Figure~\ref{FIG:OFFGRID}, where the \ac{DOA} space is sampled with a 5 degree resolution, and the ground-truth \ac{DOA} of the sources-of-interest is between 0 and 180~degree on 180 grid points (angular distance between two grid points is $\frac{180}{179}$~degree). A decreased performance of the MAE and the MedAE can be observed around 0~degree, as expected for a ULA.  Over the \ac{DOA}, the MedAE exhibits a clear triangular structure with the highest MedAE in between the sampling points of the \ac{DOA} modules. The MAE and the standard deviation,  however, do not exhibit such a clear structure. The high ACC and psACC reported in the previous experiments, the MedAE structure with the 95\% confidence interval and the standard deviation of the MAE suggest that the MAE is mostly dominated by outliers such that no clear on/off-grid structure can be observed as in the MedAE plot. In addition to SRP-MP~(SPS) and  $\text{CNN}^{\tiny{B}}_{\tiny{P}}$ with a 5~degree sampling resolution, we also evaluate SRP-MP~(SPS) with a $\frac{180}{179}$~degree sampling resolution to show the flexibility of the approach (during training, the \ac{DOA} resolution was 5~degree).  Such a change is not possible for DDNNs without the necessity to change the number of output classes of the DDNN architecture and retraining. The increased \ac{DOA} sampling resolution for SRP-MP~(SPS) leads to a reduced MedAE and MAE. This shows that scenario-dependent modifications of the \ac{DOA} module can be performed after training if the \ac{DOA} module is a signal-processing method.  

To investigate the effect of off-grid sources further, an excerpt of an estimate vs.  actual \ac{DOA} confusion plot is shown in Figure~\ref{FIG:CONFMAT}.  It is shown that for SRP-MP (SPS) and for $\text{CNN}^{\tiny{B}}_{\tiny{P}}$, the estimated \ac{DOA} of off-grid sources is typically the closest \ac{DOA} that can be estimated due to the discrete sampling resolution of 5~degrees of the estimators.  If a source \ac{DOA} is in the middle of two sampled \ac{DOA}s, the algorithms choose either of the neighboring classes.  Both algorithms show comparable performance.  Around 0~degree, as expected for a ULA, the algorithms perform worse (the \ac{DOA}-based changes in inter-microphone phase-differences have a sine dependency; the sine is zero at 0~degree, which hinders the  \ac{DOA} estimation).   

Please note that the overall results in this experiment are better compared to the previous experiments, irrespective of the off-grid evaluation due to the use of simulated RIRs here.  Additionally, SRP-MP~(SPS) performs better than the DDNN with an MAE of $3.6^\circ$  for the coarse and of $2.9^\circ$ for the fine-grid evaluation compared to an MAE of $4.6^\circ$  for the DDNN.  Physical model violations due to the use of measured RIRs with slight microphone perturbations seem to influence the performance of SRP-MP stronger than of the DDNN. Similar results have been found in \cite{Chakrabarty2017a} for SRP-P.
\subsection{Outlook: Evaluation using Measured Data of Moving Sources}
\label{CHAP:MEAS}
\begin{figure}
\centering
% This file was created by tikzplotlib v0.9.8.
\begin{tikzpicture}

\begin{axis}[
tick align=outside,
tick pos=both,
height=5cm,
width=8cm,
title={},
x grid style={white!69.0196078431373!black},
xlabel={Time [s] },
xmin=0, xmax=500,
xtick style={color=black},
xtick={0,240,500,1000,1500,2000,2500,3000},
xticklabels={{  }0,2, 4,8,12,16.6,20.75,24.9},
y grid style={white!69.0196078431373!black},
ylabel={kHz },
ymin=0, ymax=180,
ytick ={0,90,180},
yticklabels={0,4,8},
ytick style={color=black},
ytick pos=left,
xtick pos = bottom,
]
\addplot[forget plot] graphics [includegraphics cmd=\pgfimage,xmin=0, xmax=1566, ymin=0, ymax=180] {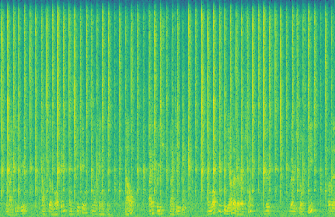};
\end{axis}
\node[below right, align=center, text=black]
at (-1, -1 ) {%
a) log-STFT of $Y$};
\end{tikzpicture}
% This file was created by tikzplotlib v0.9.8.
\begin{tikzpicture}

\begin{axis}[
tick align=outside,
tick pos=both,
height=5cm,
width=8cm,
title={ },
x grid style={white!69.0196078431373!black},
xlabel={Time [s] },
xmin=0, xmax=500,
xtick style={color=black},
xtick={0,250, 500,1000,1500,2000,2500,3000},
xticklabels={0,2, 4,8,12,16.6,20.75,24.9},
y grid style={white!69.0196078431373!black},
ylabel={kHz },
ymin=0, ymax=180,
ytick style={color=black},
ytick ={0,90,180},
yticklabels={0,4,8},
ytick pos=left,
xtick pos = bottom,
]
\addplot[forget plot] graphics [includegraphics cmd=\pgfimage,xmin=0, xmax=1566, ymin=0, ymax=180] {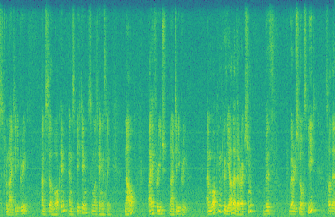};
\end{axis}
\node[below right, align=center, text=black]
at (-1, -1 ) {%
b) log-STFT of $X_1$};
\end{tikzpicture}
\caption{a) shows an excerpt of the recorded log-STFT of $Y[1,...]$ and b) shows an excerpt of the recorded log-STFT of $X_1[1,...]$. $Y$ is a superposition of separately recorded directional interference and $X$. Please note that microphone self-noise and background noise are also present in b).   The audiofile with aligned \ac{DOA} estimates is available at \url{https://www.audiolabs-erlangen.de/resources/2021-CSL-Signal-Aware-DOA-Estimation}.  }
\label{FIG:MAGSTFT}
\end{figure}
\begin{figure}
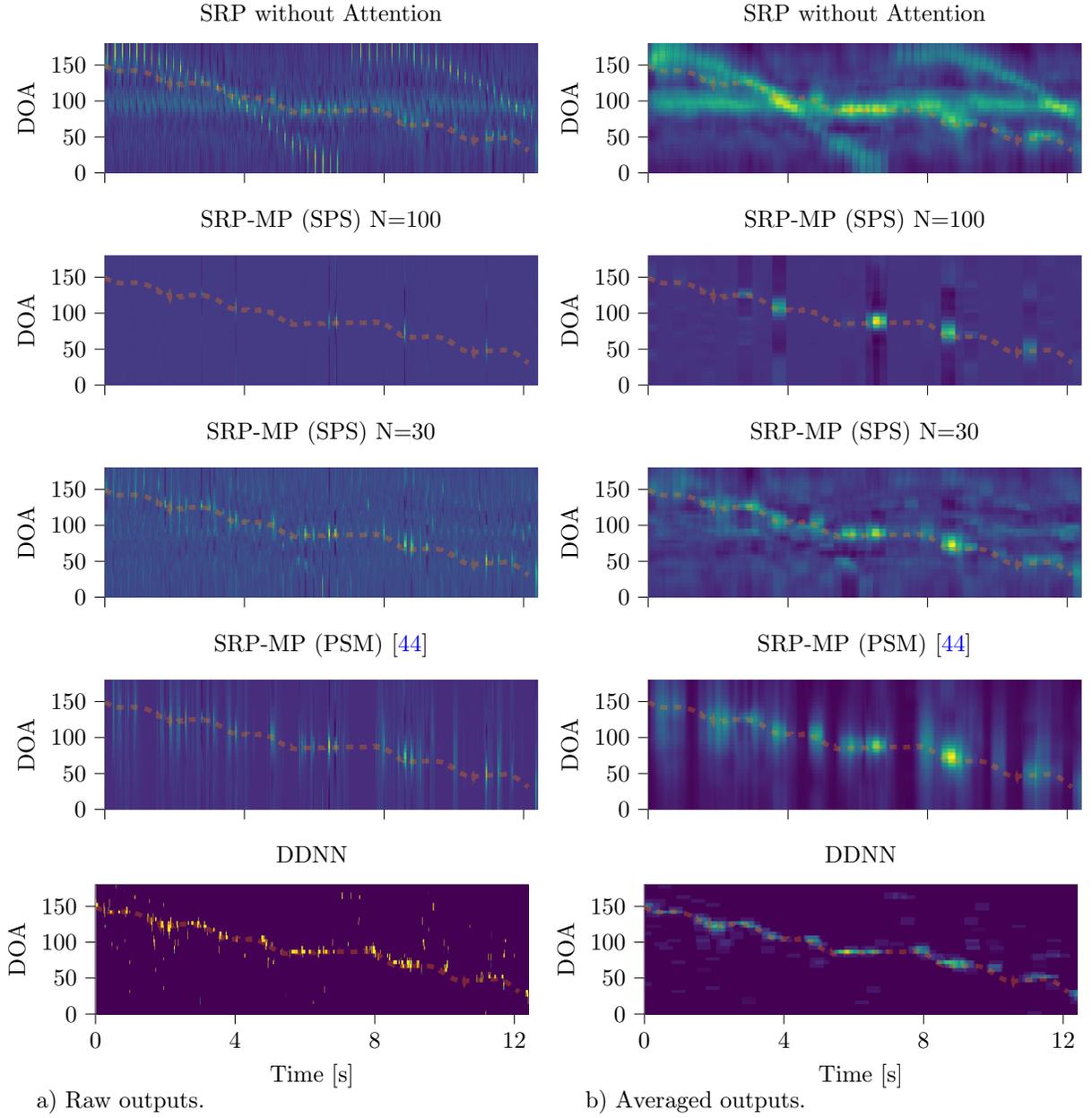

\centering
\input{SRPwithoutAttention.pdf_tex}
\input{SRPwithoutAttentionKopie.pdf_tex} \\
\input{SRP-MPE2E.pdf_tex}
\input{SRP-MPE2EKopie.pdf_tex} \\
\input{SRP-MPE2ESC.pdf_tex}
\input{SRP-MPE2ESCKopie.pdf_tex} \\
\input{SRP-MPPSM.pdf_tex}
\input{SRP-MPPSMKopie.pdf_tex} \\
\input{DDNNwithE2EAttention.pdf_tex}
\input{DDNNwithE2EAttentionKopie.pdf_tex}
\caption{\ac{DOA} estimates of measured data using the respective methods. In a), the raw outputs are shown, whereas in b) the raw outputs are averaged over 0.4~s, respectively.  There is  a single moving speech source active tracked via an OptiTrack (orange line). There are additional directional interfering sources moving from 180 to 0 degrees as can be seen in SRP without attention. The SIR was lower than $-4.5$~dB.  For SRP-MP~(SPS), the ADNN was trained with either $N=30$ or $N=100$ frames.}
\label{FIG:MEAUREMENTS}
\end{figure}

\begin{figure}
\centering
\input{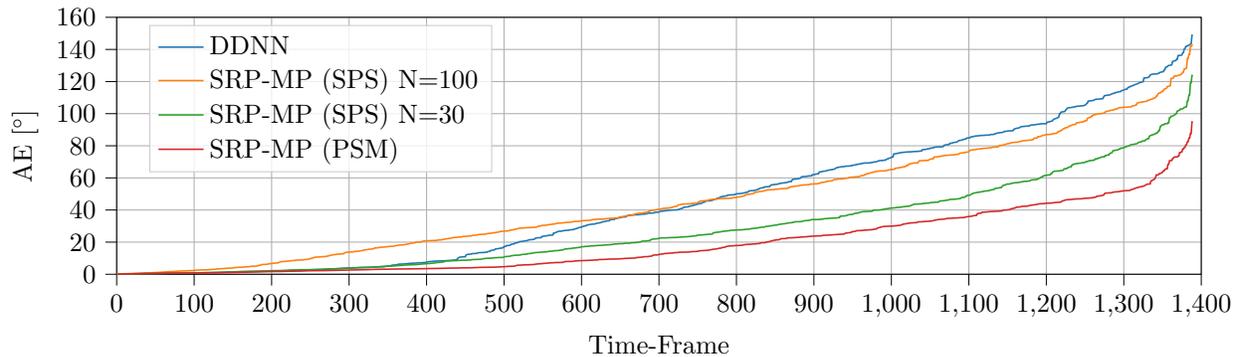}
\caption{\ac{AE} per time-frame between the estimated and the ground-truth \ac{DOA} obtained via an OptiTrack.  The figure demonstrates the number of reliable time-frames per method.  For each method, the time-frames are ordered from low AE to high AE irrespective of the output energy in the respective \ac{DOA} estimate (see Figure~\ref{FIG:MEAUREMENTS}).  Methods with very sparse outputs consequently have a very fast increasing AE curve.  There are no abrupt increases of the AE as the high number of evaluated time-frames leads to a smooth AE curve.  The SIR was smaller than $-4.5$~dB.}
\label{FIG:MEASDIFFOVERFRAME}
\end{figure}

Finally, we compare $\text{CNN}^{\tiny{B}}_{\tiny{P}}$, SRP-MP~(SPS) and SRP-MP~(PSM) using measured data for one moving speaker (the \ac{SOI}), with undesired directional sources, and diffuse background sounds with a SIR smaller than $-4.5$~dB {(room~size:~4.7~m,~4.87~m,~2.6~m~height;~$T_{60}$:~0.5~s).}  A plot of the recorded signals can be seen in Figure~\ref{FIG:MAGSTFT}. The ground-truth source positions were obtained using an optical tracking system \cite{OPTITRACKER}.  The outputs of the estimators are visualized in Figure~\ref{FIG:MEAUREMENTS} (not normalized). The output of SRP-P shows a stationary directional interfering source at approximately $95{^\circ}$ and moving directional interfering sources.  The \ac{DOA} of the speech source is barely visible.  Such an output is expected if no attention is applied.  The output of SRP-MP~(SPS) is temporally very sparse, however, the estimates are very close to the ground truth obtained via optical tracking. This result is consistent with the sparse masks of SRP-MP~(SPS) seen in Figure~\ref{Fig:SRP}. In this low-SIR environment, SRP-MP~(SPS) only bases the \ac{DOA} estimates on very few time-frames that yield very accurate estimates. This effect can be explained via the objective function (\ref{EQU:LOSSE2ESRP}), where the estimated \ac{SPS} is optimized to resemble the \ac{SPS} of the ground-truth, both after normalization (see (\ref{EQU:SRPnormalize})).  The normalization removes the energy-diminishing effects of the mask/attention such that a very sparse mask can still yield a very low loss.  Such sparse estimates are acceptable for static or slowly moving sources, however,  not for tracking moving sources or a very short temporal context.  

An additional experiment was conducted to study the influence of the length $N $of the training context window on the output. We expect that the outputs are less sparse if SRP-MP~(SPS) is trained with shorter context windows ($N=30$ instead of $N=100$ time-frames, during training), denoted as SRP-MP~(SPS)~$N=30$.  As shown in Figure~\ref{FIG:MEAUREMENTS},  the outputs of SRP-MP~(SPS)~$N=30$ are less sparse than of SRP-MP~(SPS) at the cost of reduced performance on {\scshape{test-2S}} (79~\%,  67~\%, 8.7~$^\circ$, for psACC, ACC, MAE, respectively).  SRP-MP~(SPS) yields sharper outputs than SRP-MP~(PSM) in Figure~\ref{FIG:MEAUREMENTS}. However, SRP-MP~(PSM) also yields less sparse outputs than SRP-MP~(SPS).  The outputs are temporally less sparse as the ADNN was trained to resemble the PSM that is defined on a time-frequency bin-basis.  Less-sparse outputs are advantageous for tracking. Interestingly,  as shown in Table~\ref{TAB:RESE2EADNN}, the sharper outputs of SRP-MP~(SPS) improve the ACC only by 2~\% compared to SRP-MP~(PSM). The DDNN yields very narrow non-sparse estimates that are very close to the path of the speech source.  In contrast to SRP-MP~(SPS), the outputs are not sparse as the ADNN objective in (\ref{EQU:E2ELOSSFM}) does not exhibit a normalization of the estimates. The non-sparse outputs make the DDNN  more suited for tracking than SRP-MP~(SPS). 

All methods also generalize to measured data under low SIR/SNR conditions and moving sources. The outputs of SRP-MP~(PSM) are much broader than those of the DDNN. We compare the  frame-wise \ac{AE} (reordered time-frames) of the respective file in Figure~\ref{FIG:MEASDIFFOVERFRAME} to investigate whether this is a visualization issue only.  As expected, SRP-MP~(SPS) and SRP-MP~(SPS)~$N=30$  exhibit very few frames with a low error as the estimated outputs are temporally very sparse. SRP-MP~(SPS)~$N=30$  has a lower AE for more time-frames, as expected. The DDNN and SRP-MP~(PSM) perform comparably for approximately 400 frames before the AE of the DDNN begins to increase rapidly. Consequently, the broad lobes of SRP-MP~(PSM) are misleading in terms of frame-wise MAE for a single \ac{SOI}. As multi-source localization of sparse sources (like speech) can be broken down to single-source localization, the broad lobes of SRP-MP~(SPS) pose no disadvantage for localization. Figure~\ref{FIG:MEAUREMENTS} shows that both the DDNN and SRP-MP~(PSM) are suited for tracking as there are temporally non-sparse estimates that follow the \ac{SOI} path. 

As all methods estimate the \ac{DOA} on an \ac{STFT} time-frame basis, an extension to more sophisticated source tracking can be achieved,  for example,  by using recurrent neural networks or particle filters that process the frame-wise \ac{DOA} estimates of the proposed and baseline methods (see, e.g., \cite{sharath2019}).  For tracking,  temporally non-sparse estimates are beneficial if a high temporal \ac{DOA} resolution is required; for this, SRP-MP~(SPS) is less suited than SRP-MP~(PSM). Finally, the DDNN and SRP-MP~(PSM) both seem to be suited for tracking. As SRP-MP~(PSM) has an overall lower computational complexity than the DDNN and performs comparable in all tests, we conclude that hybrid approaches represent a low complexity, high flexibility, highly accurate method for signal-aware \ac{DOA} estimation.  
\acresetall
\section{Conclusion}
We used a \ac{DNN} to estimate attention from a single-channel microphone spectrum. The attention was subsequently used in a fully \ac{DNN}-based system or a hybrid fashion in signal processing-based methods for signal-aware \ac{DOA} estimation of speech sources.  We showed that spectral context is crucial for \ac{DNN}-based \ac{DOA} estimators and that they are biased towards the source classes and the attention distribution seen during training. In contrast, signal processing-based \ac{DOA} estimation does not exhibit such a bias. We proposed \ac{DOA}-based training objectives for fully data-driven and hybrid signal-aware \ac{DOA} estimators and showed that both variants perform comparably.  We also showed that the spectrum of a single microphone is sufficient to compute attention, making the attention computation independent of the array architecture, assuming a signal-processing method is used for subsequent \ac{DOA} estimation.  This is especially interesting as, in contrast to \ac{DNN}s,  signal-processing methods can be modified and adapted during runtime without retraining the attention estimator to match different conditions,  like array architectures,  inter-microphone distances, number of microphones,  etc.   We conclude that hybrid systems pose a low complexity, high flexibility approach for signal-aware \ac{DOA} estimation with comparable performance to fully \ac{DNN}-based signal-aware \ac{DOA} estimation.
\section*{Acknowledgment}
The authors would like to thank the Erlangen Regional Computing Center (RRZE) for providing computing resources and support. We want to thank Anna Leschanowsky for assisting in recording optical tracking data.
\bibliography{SigAwareDOA}
\end{document}